\documentclass[10pt,twocolumn,showpacs,preprintnumbers,amsmath,amssymb,aps,prl,superscriptaddress,longbibliography]{revtex4-2}
\usepackage{appendix}
\usepackage{bbm}
\usepackage{mathrsfs}
\usepackage{graphicx}
\usepackage{dcolumn}
\usepackage{bm}
\usepackage{braket}
\usepackage{amsmath}
\usepackage{amsfonts}
\usepackage{CJKutf8}
\usepackage[dvipsnames]{xcolor}
\usepackage[colorlinks=true,linkcolor=magenta,urlcolor=BlueViolet,citecolor=BlueViolet]{hyperref}
\usepackage{natbib}
\usepackage{multirow}

\begin{document}

\title{Orbital Description of Landau Levels}
\author{Huan Wang}
\affiliation{State Key Laboratory of Surface Physics and Department of Physics, Fudan University, Shanghai 200433, China}
\affiliation{Shanghai Research Center for Quantum Sciences, Shanghai 201315, China}
\author{Rui Shi}
\affiliation{State Key Laboratory of Surface Physics and Department of Physics, Fudan University, Shanghai 200433, China}
\affiliation{Shanghai Research Center for Quantum Sciences, Shanghai 201315, China}
\author{Zhaochen Liu}
\affiliation{State Key Laboratory of Surface Physics and Department of Physics, Fudan University, Shanghai 200433, China}
\affiliation{Shanghai Research Center for Quantum Sciences, Shanghai 201315, China}
\author{Jing Wang}
\thanks{Contact author: wjingphys@fudan.edu.cn}
\affiliation{State Key Laboratory of Surface Physics and Department of Physics, Fudan University, Shanghai 200433, China}
\affiliation{Shanghai Research Center for Quantum Sciences, Shanghai 201315, China}
\affiliation{Institute for Nanoelectronic Devices and Quantum Computing, Fudan University, Shanghai 200433, China}
\affiliation{Hefei National Laboratory, Hefei 230088, China}

\begin{abstract}
The pursuit of a lattice analogue for Landau levels has been a central theme in condensed matter physics. Although the correspondence between Chern bands and the lowest Landau level has been widely studied, a lattice realization of the first Landau level remains elusive. Here we construct a minimal lattice model that provides a concrete orbital description of both the lowest and first Landau levels. Using maximally localized Wannier functions with $s$, $p_-$, and $p_+$ orbital character, we develop a three-orbital model in which the two lowest Chern bands are flat and each carries a Chern number $\mathcal{C}=1$. The band topology arises from a sequence of ideal band inversions between Wannier states at the $\Gamma$ and $K$ points in momentum space, establishing an adiabatic connection between the atomic insulator limit and Landau level physics. Notably, many-body exact diagonalization reveals that the non-Abelian state can appear in the half-filled first Chern band. This construction can be further generalized to realize flat Chern bands analogous to higher Landau levels. Our results provide a new perspective on lattice analogues of Landau levels and may enable the exploration of fascinating topological phenomena at elevated temperatures.
\end{abstract}

\date{\today}

\maketitle

\emph{Introduction---}The recent discovery of Abelian fractional Chern insulators (FCI) at zero magnetic field, dubbed the fractional quantum anomalous Hall state in moir\'e MoTe$_2$ and pentalayer graphene~\cite{cai2023,zeng2023,park2023,xu2023,lu2024a} has generated intense interest in this new state of matter~\cite{li2021,crepel2023,cano2023,yu2024,dong2023a,goldman2023,wang2024a,jia2024,macdonald2024,ji2024,redekop2024,kang2024,xie2024,lu2024b}. The emergence of fractional topological states is attributed to the existence of flat Chern bands~\cite{tang2011,sun2011,neupert2011,qi2011,sheng2011,regnault2011,Shi2024PRR} with nearly ideal quantum geometry in moir\'e superlattice~\cite{Roy2014,mera2021, wang2021,Ledwith2020,ledwith2023}, which resembles the lowest Landau level (LLL). The exotic non-Abelian topological orders~\cite{willet1987}, such as Moore-Read states~\cite{moore1991,read2000,bonderson2006,levin2007} or Read-Rezayi state~\cite{read1999} are predicted to exist in the partially filled first Landau level (1LL). These states support non-Abelian quasiparticle excitations~\cite{read2000} and could be utilized as a platform for fault-tolerant quantum computation~\cite{kitaev2003,nayak2008}. This raises an interesting question~\cite{fujimoto2024,reddy2024,xu2024,ahn2024,chen2024,wang2024b}: Is it possible to construct a lattice analogy of 1LL without explicitly invoking the magnetic field? This question is of significant importance because such a construction could pave the way for realizing non-Abelian fractionalization at elevated temperatures. 

Here, rather than relying on the operator algebra to bridge Chern bands and LLs, we establish this connection via an adiabatic pathway from atomic limits to LL physics within the skyrmion Chern band model. We construct a projected three-band lattice model using maximally localized Wannier functions (MLWFs) with $s$, $p_-$, and $p_+$ orbital characteristics. The LL-like band topology in our model emerges through a sequence of ideal band inversions at high-symmetry points. We show using many-body exact diagonalization that the half-filled first Chern band (1CB) supports non-Abelian Moore-Read state. Our approach can be further extended to engineer flat Chern bands that mimic higher LLs. Notably, the corresponding lattice model is well-suited for implementation in cold atom systems, where the long-range Coulomb interaction is effectively projected onto a short-range density–density interaction.

\emph{Intuition---}As a starting point for constructing Chern bands in a lattice model, it is essential to understand the characteristics of the corresponding Wannier states. In earlier work, Qi constructed the one-dimensional MLWF in Chern bands that have a one-to-one mapping to the LLL wave functions in quantum Hall (QH)~\cite{qi2011} of the form $\psi_{0,K_y}(x,y)\sim e^{iK_yy}H_0(x')e^{-x'^2/2l^2_B}$, where $x'\equiv x-K_y\ell_B^2$. $H_0(x')$ is the Hermite polynomial for $n=0$ LLL, and $H_0(x')e^{-x'^2/2l^2_B}=e^{-x'^2/2l^2_B}$ is related to MLWF of Chern band and exhibits $s$-orbital–like character in one dimension. Intuitively, when this construction is extended to the 1LL, the MLWFs in the corresponding Chern band map to $H_1(x')e^{-x'^2/2l^2_B}=2x'e^{-x'^2/2l^2_B}$, which has a $p$-orbital–like structure. This suggests that, at a minimum, Wannier states with $s$- and $p$-orbital characteristics are necessary to construct a Chern band that captures the physics of the 1LL.

\emph{Model and adiabatic connection---}We begin with the skyrmion Chern band model in which itinerant electrons couple to a layer pseudospin skyrmion lattice. This model could well describe the consecutive topologically nontrivial flat bands in twisted homobilayer MoTe$_2$ around twist angle $2^\circ$~\cite{reddy2024,wu2019,wang2024b,devakul2021,reddy2023,zhang2024polarization,wu2024,shi2024}, 
\begin{equation}
    \mathcal{H}_0=\frac{\mathbf{p}^2}{2m}+\mathcal{J}\bm{\sigma}\cdot\mathbf{S}(\mathbf{r}).
    \label{eq1}
\end{equation}
Here, $\bm{\sigma}$ denotes the pseudospin operator associated with the layer degree of freedom, and $\mathbf{S}(\mathbf{r})=\mathbf{S}(\mathbf{r}+\mathbf{a}_{1,2})$ represents the periodic moir\'e potential and is naturally coupled to the pseudospin of the layer with strength $\mathcal{J}$. In the small twist-angle limit, the moir\'e potential becomes dominant, such that $\mathcal{J} \gg \hbar^2/ma^2$, and enforces the local alignment of the electron layer polarization with $\mathbf{S}(\mathbf{r})$. This alignment induces a pseudospin Berry phase. The effect can be made explicit through a position-dependent $SU(2)$ unitary transformation $\mathbf{U}(\mathbf{r})$, which rotates the pseudospin texture $\mathbf{S}(\mathbf{r})$ into $S(\mathbf{r})\hat{z}$, thereby introducing a gauge field $\mathbf{A}_j=(i\hbar/e)\mathbf{U}^\dagger \partial_j \mathbf{U}$. In the large-$\mathcal{J}$ limit, the system can be projected onto the low-energy manifold of locally layer-polarized electrons, yielding an effective Hamiltonian for the low-energy band as
\begin{equation}
    \mathcal{H}_{\text{eff}}=\frac{\left(\mathbf{p}-e\mathbf{A}(\mathbf{r})\right)^2}{2m}+\sum\limits_{j=x,y}\frac{\hbar^2}{8m}(\partial_j \hat{\mathbf{S}})^2-\mathcal{J}S(\mathbf{r}),
    \label{eq2}
\end{equation}
where $\mathbf{A}(\mathbf{r})$ is the $\downdownarrows$ component of the emergent $SU(2)$ gauge field, with $\nabla\times\mathbf{A}(\mathbf{r})\equiv B^e(\mathbf{r})=(\hbar/2e)\hat{\mathbf{S}}\cdot(\partial_x \hat{\mathbf{S}} \times \partial_y \hat{\mathbf{S}})$ be spatially non-uniform. $\mathcal{H}_{\text{eff}}$ 
describes a pseudo-spinless electron in a magnetic field when the last term in Eq.~(\ref{eq2}) is a constant, where the energy bands form dispersive LL. However, the potential $\mathbf{S}(\mathbf{r})$ fluctuates (sometimes wildly) according to the different local layer structures in the moir\'e system, so $S(\mathbf{r})$ is no longer generically constant. For large $\mathcal{J}$, the last term in Eq.~(\ref{eq2}) may dominate and its minimum forms a potential trap. Each trap has the form as $-\mathcal{J}S(\mathbf{r})\sim (\mathbf{r}-\mathbf{r}_{\text{min}})^2 + O(\mathbf{r}-\mathbf{r}_{\text{min}})^3$, which respects the rotational symmetry around the minimum, thus $s$, $p$, $d$ type orbitals emerge in the low-energy bands~\cite{kariyado2019,angeli2021,liu2022}. Thus, in this limit, $\mathcal{H}_{\text{eff}}$ describes orbital hopping with effective magnetic flux.

\begin{figure}[t]
    \begin{center}
    \includegraphics[width=3.4in,clip=true]{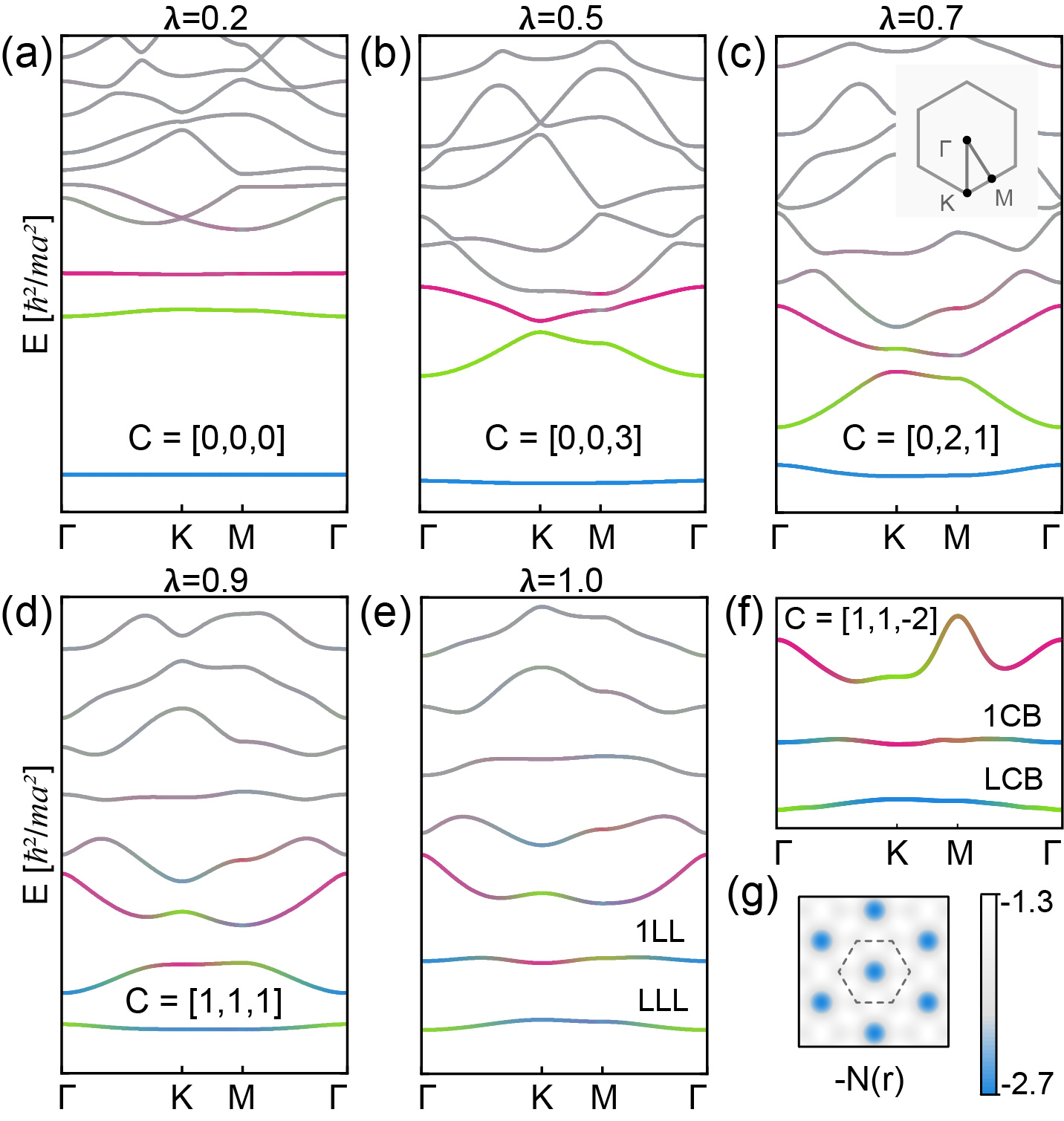}
    \end{center} 
    \caption{Band structures and adiabatic connection from atomic insulator to LL. (a)-(e) The lowest few bands of $\mathcal{H}_0$ with different normalization parameter $\lambda$, where the overlap between Bloch state and Wannier function is labeled as different colors, blue for $s$ orbital, green for $p_-$ orbital, red for $p_+$ orbital and $\mathcal{C}$ denotes the Chern number of lowest three bands. The inset in (c) shows the Brillouin zone (BZ). (f) Band structure of the minimal three-orbital tight-binding model. (g) Real space distribution of $-N(\mathbf{r})$. The parameters are $\mathcal{J}/(\hbar^2/ma^2)=52\pi^2$, $\alpha=1$ and $N_0=0.28$.}
    \label{fig1}
\end{figure}

From such observations, we see that $\mathcal{H}_{\text{eff}}$ actually links effective atomic orbital and LL physics in the low-energy manifold. This is made clearly by calculating the band structure of $\mathcal{H}_0$ when we take $\mathbf{S}(\mathbf{r})=\mathbf{N}(\mathbf{r})/N^\lambda(\mathbf{r})$ with
\begin{equation}
    \mathbf{N}(\mathbf{r})=\frac{1}{\sqrt{2}}\sum_{j=1}^{6}e^{i\mathbf{q}_j\cdot\mathbf{r}}\hat{\mathbf{e}}_j+N_0\hat{z},
    \label{eq3}
\end{equation}
where $\hat{\mathbf{e}}_j=\left(i\alpha\sin\theta_j,-i\alpha\cos\theta_j,-1\right)/\sqrt{2}$, $\mathbf{q}_j=(4\pi/\sqrt{3}a)\left(\cos\theta_j,\sin\theta_j\right)$, and the angles satisfy $\theta_2=\theta_1+2\pi/3$, $\theta_3=\theta_1+4\pi/3$, and $\theta_{j+3}=\theta_{j}+\pi$. This pseudospin texture can be viewed as a sum of three pseudospin spirals forming a triangular skyrmion lattice~\cite{tonomura2012,karube2017,tokura2021,lin2016}, and the normalization is controlled by $\lambda$. When $\lambda=1$, $S(\mathbf{r})$ becomes spatially uniform, and the low energy bands of $\mathcal{H}_0$ are flat Chern bands with $\mathcal{C}=1$, as shown in Fig.~\ref{fig1}(e). The lowest two Chern bands are dispersive LL, where their wave functions overlap with the corresponding flat LL exceed $99\%$~\cite{supple}. For simplicity, we label them as LLL and 1LL, respectively. Conversely, when $\lambda \rightarrow 0$, the minimum of $-\mathcal{J}S(\mathbf{r})$ forms a moir\'e triangular lattice (see Fig.~\ref{fig1}(g)), and three lowest bands in Fig.~\ref{fig1}(a) are topologically trivial flat bands resembling the atomic insulator limit, which emerge from $s$, $p_-$, $p_+$ orbital hopping on a triangular lattice. 

Fig.~\ref{fig1} display the evolution of the band structures from flat topologically trivial bands to flat Chern bands. As $\lambda$ changes from 0 to 1, the lowest three bands invert while their $s$, $p_-$, $p_+$ orbital characteristics remain. To see the band inversion clearly, we graphically depict the overlap of the Bloch states with trial Wannier functions of $s$, $p_-$, and $p_+$ orbitals, symbolized by blue, green, and red, respectively. The lowest three bands first invert with the upper band at $M$ and their Chern numbers become $\mathcal{C}=[0,0,3]$ as shown in Fig.~\ref{fig1}(b), then the second and third lowest bands invert at $K$ and switch Chern numbers to $\mathcal{C}=[0,2,1]$ as in Fig.~\ref{fig1}(c), finally the lowest and second lowest bands invert at $\Gamma$ and Chern numbers becomes $\mathcal{C}=[1,1,1]$ as in Fig.~\ref{fig1}(d,e). The lowest three Chern bands are isolated in Fig.~\ref{fig1}(e). The orbital characteristics of the LLL and 1LL here are consistent with the mapping from the hybrid Wannier functions of Chern band to LL wave functions in QH~\cite{qi2011}.

\begin{figure}[t]
    \begin{center}
    \includegraphics[width=3.0in,clip=true]{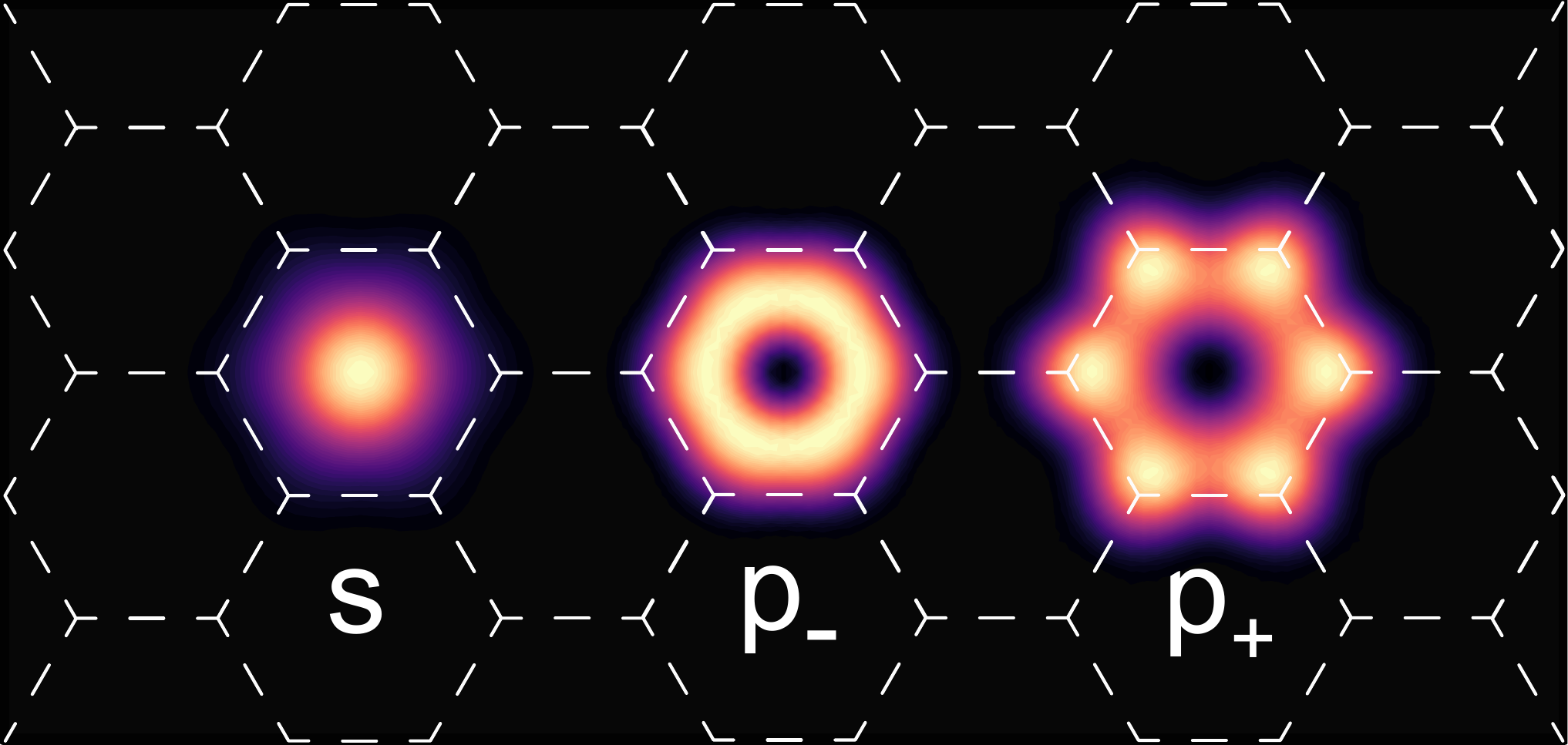}
    \end{center} 
    \caption{The MLWFs of the lowest few bands in Fig.~\ref{fig1}(e) with $s$, $p_-$, $p_+$ orbital characteristics, and the white hexagon labels Wigner-Seitz cell.}
    \label{fig2}
\end{figure}

\emph{Wannier projection---}Now we demonstrate that $s$, $p_-$, $p_+$ orbitals constitute a complete subspace to describe LLL and 1LL. By constructing the MLWFs~\cite{pizzi2020,sakuma2013,marzari1997,marzari2012} from the continuum model, a three-orbital model is developed to describe the lowest two Chern bands in Fig.~\ref{fig1}(e), with the Wannier orbitals forming a triangular lattice. The minimal model includes hopping up to the fifteenth nearest neighbor, where its explicit form is in Supplemental Materials~\cite{supple}. The band structure in Fig.~\ref{fig1}(e) does not possess a local gap below which the total Chern number is zero. A set of frozen states is chosen to preserve the topology of the focused Chern bands and the band disentanglement~\cite{souza2001} process is then performed to avoid Wannier obstruction~\cite{brouder2007}. We choose the lowest two Chern bands as frozen states, and get the minimal tight-binding model where only MLWFs with $s$, $p_-$, $p_+$ orbital characteristics are included. The corresponding band structure is calculated in Fig.~\ref{fig1}(f). The MLWFs of $s$, $p_-$, $p_+$ orbitals are shown in Fig.~\ref{fig2}. The lowest two bands projected from the LLL and 1LL are well reconstructed in Fig.~\ref{fig1}(f), which are labeled LCB (lowest Chern band) and 1CB, respectively.

\begin{figure}[t]
    \begin{center}
    \includegraphics[width=3.4in,clip=true]{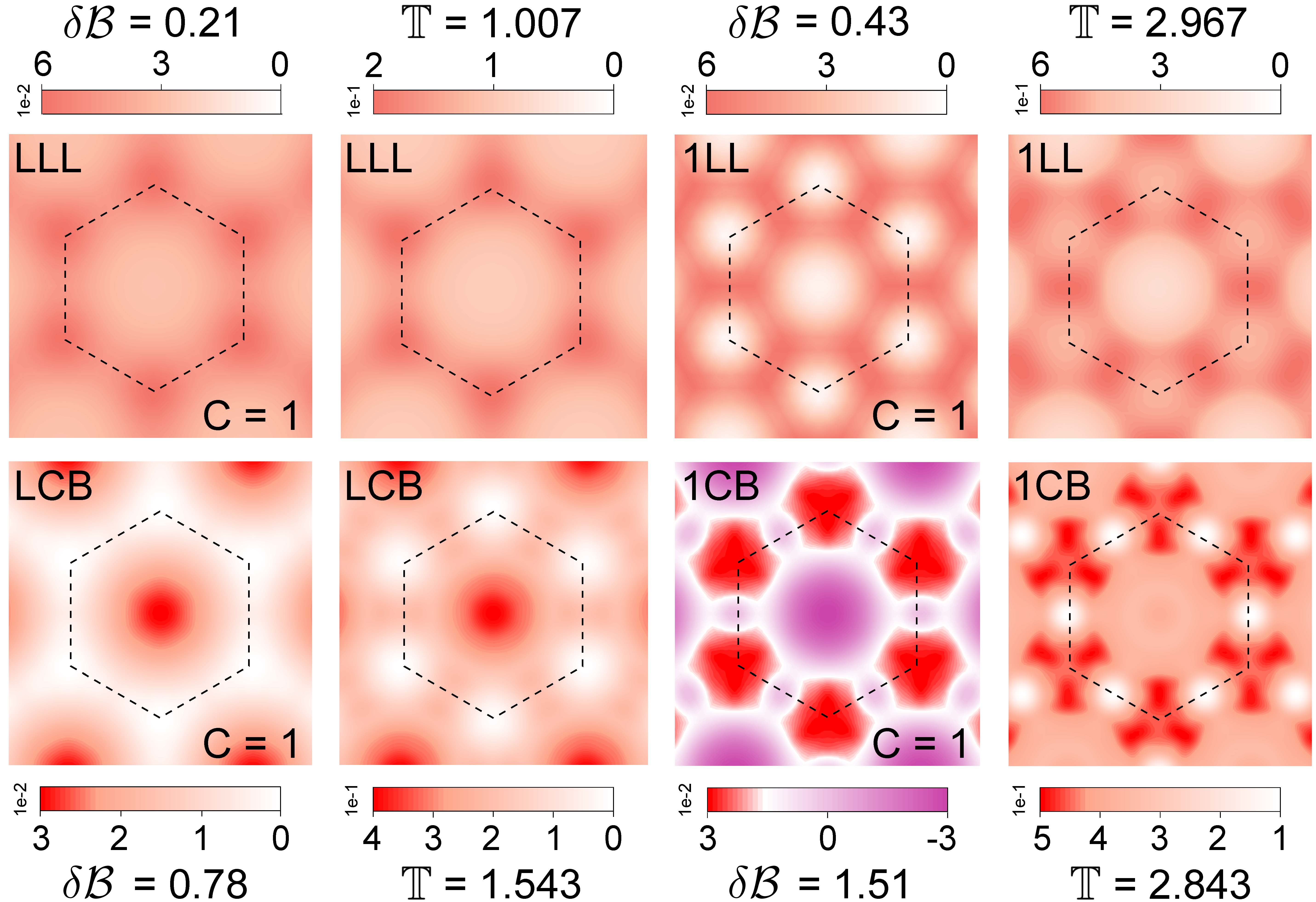}
    \end{center} 
    \caption{Distribution of Berry curvature $\mathcal{B}(\mathbf{k})$ and trace of Fubini–Study metric $\text{Tr}[g(\mathbf{k})]$ for the two lowest bands of both continuum model and tight-binding model across the Brillouin zone (BZ). The BZ is outlined by dashed hexagon. The bands derived from the continuum model are designated as LLL and 1LL, reflecting their highly ideal character and close resemblance to flat LLs. In contrast, the bands from the tight-binding model are labeled LCB and 1CB. The Berry curvature fluctuation $\delta\mathcal{B}$ is noticeably larger in LCB and 1CB compared to their LLL and 1LL counterparts, resulting in a stronger deviation of the trace condition measure $\mathbb{T}$ from the ideal value.}
    \label{fig3}
\end{figure}

Two band geometry indicators are employed to evaluate the different between Chern bands and LL~\cite{Parameswaran2012,Roy2014,jackson2015,claassen2015,mera2021,ozawa2021,wang2021,ledwith2023}, namely Berry curvature fluctuation $\delta\mathcal{B}$ and average trace condition $\mathbb{T}$ (non-negative) defined as 
\begin{align}
    (\delta\mathcal{B})^2 &\equiv \frac{\Omega_{\text{BZ}}}{4\pi^2}\int_{\text{BZ}} d\mathbf{k} \left(\mathcal{B}(\mathbf{k})-\frac{2\pi \mathcal{C}}{\Omega_{\text{BZ}}}\right)^2, \\
    \mathbb{T} &\equiv \int_{\text{BZ}} d\mathbf{k} \left[\mathrm{Tr}(g(\mathbf{k}))\right],
\end{align}
where $\mathcal{B}(\mathbf{k})\equiv-2\mathrm{Im}(\eta^{xy})$ is the Berry curvature, $g(\mathbf{k})\equiv\mathrm{Re}(\eta^{\mu\nu})$ is the Fubini-Study metric, and $\eta^{\mu\nu}(\mathbf{k})\equiv\langle \partial^\mu u_{\mathbf{k}}| \left(1- |u_{\mathbf{k}}\rangle\langle  u_{\mathbf{k}}|\right) |\partial^\nu u_{\mathbf{k}}\rangle$ is the quantum geometric tensor, $\mathcal{C}\equiv(1/2\pi)\int d^2\mathbf{k}\mathcal{B}(\mathbf{k})$, $\Omega_{\text{BZ}}$ is area of BZ. We plot the distribution of $\mathcal{B}(\mathbf{k})$ and $\text{Tr}[g(\mathbf{k})]$ of the BZ in Fig.~\ref{fig3}. For LLL and 1LL, $\mathcal{B}(\mathbf{k})$ remain positive throughout the whole BZ and the distribution is quite homogeneous with relatively small fluctuation $\delta \mathcal{B}$, and $\mathbb{T}$ is almost ideal ($\mathbb{T}=2n+1$ for $n$th flat LL~\cite{liu2025}). However, in the tight-binding model, $\mathcal{B}(\mathbf{k})$ is no longer homogenous but concentrates around the band inversion points, and their sign is not always positive (such as in 1CB) throughout the BZ, leading to relatively large fluctuation $\delta\mathcal{B}$ and deviation of $\mathbb{T}$ from the ideal value in LL. 

These differences between Chern bands of the tight-binding model and dispersive LL of the continuum model are inevitable, since we project out the higher energy degree of freedom in the continuum model and keep only $s$, $p_-$, $p_+$ Wannier states. Then the band inversion among them naturally lead to Berry curvature concentration. Especially for the 1CB, the band inversion between $p_-$ and $s$ contributes the negative $\mathcal{B}(\mathbf{k})$ around $\Gamma$ point and the inversion between $p_-$ and $p_+$ contribute the positive $\mathcal{B}(\mathbf{k})$ around $K$ and $K^\prime$ points. Remarkably, we will see that a non-Abelian state can occur even when the Berry curvature fluctuates strongly.

\emph{Exact diagonalization---}To explore whether the non-Abelian state can appear in 1CB, we now study many-body physics at fractional filling in these Chern bands via numerical diagonalization. To make the many-body calculation tractable, we restrict our variational Hilbert space to that in which $N_{\text{uc}}$ electrons fill the LCB and $N_e-N_{\text{uc}}$ electrons remain in the 1CB where $N_e$ is the number of electrons. The electron-electron interaction Hamiltonian is obtained by projecting the realistic Coulomb interaction into the MLWFs and keeping the leading terms~\cite{supple}, which is defined as
\begin{equation}
    \mathcal{H}_{\text{int}}= \frac{U}{2} \sum_{i,a\neq b} n_{i,a}n_{i,b} + V \sum_{\left<ij\right>,a,b} n_{i,a}n_{j,b},
\end{equation}
where $a$,$b$ = ($s$, $p_-$, $p_+$), $\left<ij\right>$ means nearest-neighbor, $U$ and $V$ are the strength of onsite and nearest-neighbor interaction. We adopt an isotropic approximation, neglecting minor variations in the interactions between different orbital pairs~\cite{supple}. We also neglect the kinetic energy since these two lowest Chern bands are quite flat.

\begin{figure}[t]
    \begin{center}
    \includegraphics[width=3.4in,clip=true]{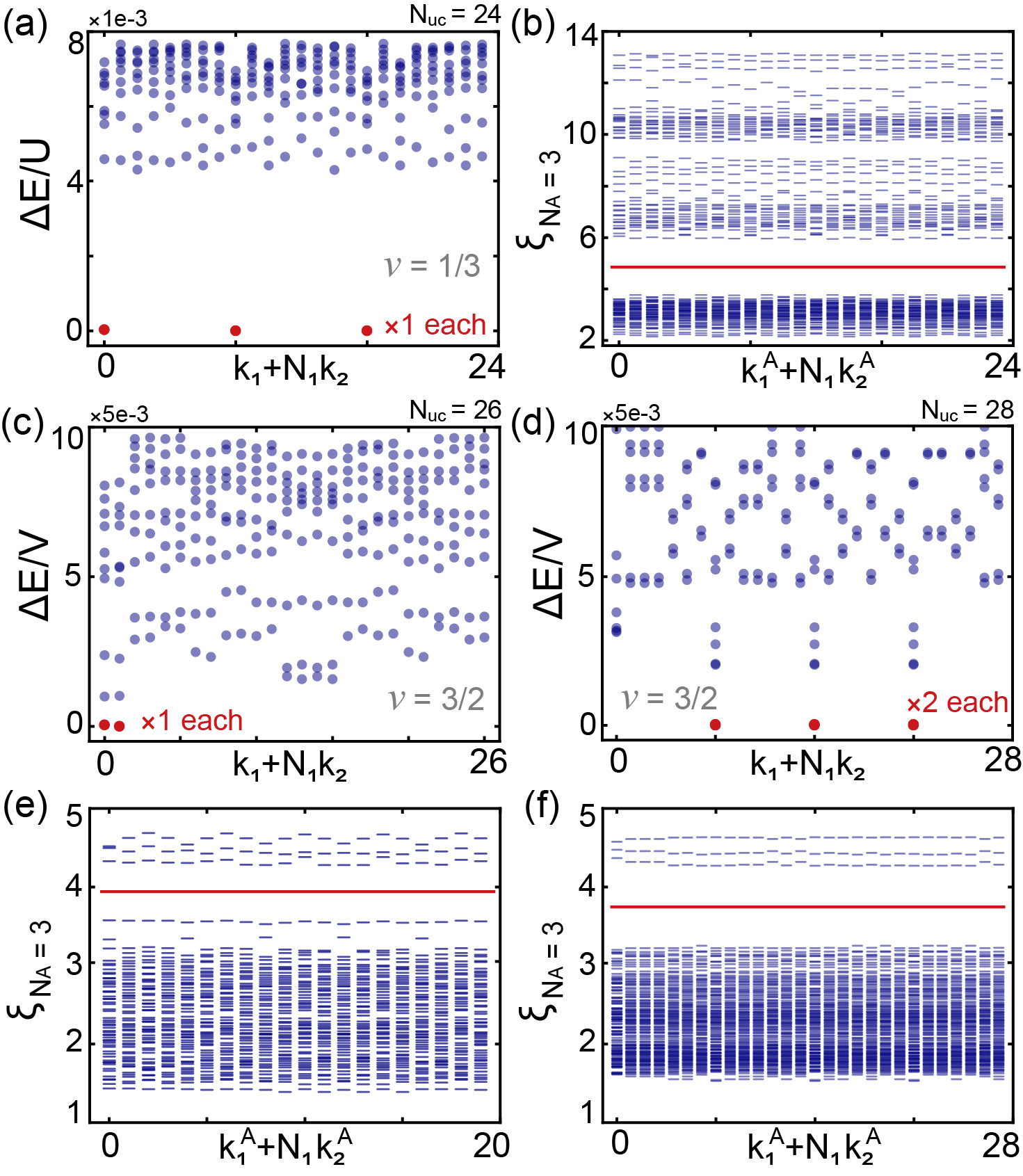}
    \end{center} 
    \caption{Exact diagonalization and PES for fractional filled CB. (a) Low energy many-body energy spectrum for $1/3$ filled LCB with $N_{\text{uc}}=24$. (b) PES with $N_A=3$ for the three degenerate ground states in (a), where $\rho=(1/d)\sum_{i}^{d}\left|\Psi_i\right>\left<\Psi_i\right|$. $d$ is the degeneracy of ground states. (c), (d) Energy spectrum for $1/2$ filled 1CB with $N_{\text{uc}}=26$ and $N_{\text{uc}}=28$. (e), (f) PES with $N_A=3$ for the six degenerate ground states in $N_{\text{uc}}=20$ and $N_{\text{uc}}=28$. Here we only show the lowest energy per momentum sectors in addition to the degenerate ground state.}
    \label{fig4}
\end{figure}

Fig.~\ref{fig4}(a) displays the many-body spectra at filling $\nu=1/3$ (that is, $1/3$ filling of LCB) as a function of crystal momentum $\mathbf{k}=k_1\mathbf{T}_1 + k_1\mathbf{T}_2$,  which is labeled $k=k_1 + N_1k_2$. Here $k_{i}=0,...,N_{i}-1$ ($i=1,2$) for system size $N_{\text{uc}}=N_1\times N_2$ with filled particle number $N_e=\nu N_{\text{uc}}$ and $\mathbf{T}_i$ are basis vectors of crystal momentum. The cluster size is chosen as $N_{\text{uc}}=4\times6$, and $U=1$ and $V=0$. There are 3 nearly degenerate ground states well separated by a sizable energy gap from excited states. The approximate ground state degeneracy matches with the expected topological degeneracy of a fractional QH state on a torus. We further calculated different cluster sizes with all other parameters fixed and found that the gap remains, indicating its existence in the thermodynamic limit~\cite{supple}. The lattice momenta of the degenerate ground states have linear indices $(0,8,16)$ and are in precise agreement with the generalized Pauli principle, which is the hallmark of FCI at $1/3$ filling~\cite{regnault2011}. To further confirm and distinguish FCI and other competing phases, we subsequently calculated the particle entanglement spectrum (PES) which encodes the information of the quasihole excitations~\cite{sterdyniak2011,chandran2011}, by dividing the whole system into $N_A=3$ and $N_e-N_A$ particles. As shown in Fig.~\ref{fig4}(b), we find that there is a clear entanglement gap separating the low-lying PES levels from higher ones for degenerate many body ground states. The number of PES levels below the gap exactly matches the typical counting of quasiparticle excitations resulting from the generalized Pauli principle of $1/3$ Laughlin state. These numerical results suggest that LCB resembles LLL.

Fig.~\ref{fig4}(c,d) show the many-body spectra at filling $\nu=3/2$ (that is, $1/2$ filling of 1CB assuming full spin polarization). We assume the energy separation between LCB and 1CB is much larger than the interaction ($U, V$), which prevents electrons in the filled LCB from gaining enough energy to be excited and participate in the many-body interactions involving the partially filled 1CB. Under these assumptions, the exact diagonalization is performed on finite-size torus with $U=1.3$ and $V=1$. For $N_{\text{uc}}=2\times13$, the number of electrons occupying the 1CB is odd (13), while for $N_{\text{uc}}=2\times14$, it is even (14). In these two cases, we observe two-fold and sixfold ground state quasi-degeneracies in Fig.~\ref{fig4}(c) and~\ref{fig4}(d), respectively. These are precisely the degeneracies expected for a Moore-Read state on the torus due to an even-odd effect~\cite{oshikawa2007,papic2012,read2000}, and the enhancement of gap indicates its existence in the thermodynamic limit. The lattice momenta at which these ground states occur also match the momenta of non-Abelian $\nu$ = 1/2 FCI based on the fractional QH-FCI folding scheme~\cite{regnault2011,Bernevig2012}. We also find that under the twist boundary condition, the quasi-degenerate ground states remain well separated from the other low-energy excitation spectrum, indicating the robustness of the excitation gap~\cite{supple}.

To exclude other competing phases in the half-filled 1CB, we further calculate PES with $N_A=3$ for $N_{\text{uc}}=4\times5$  and $N_{\text{uc}}=2\times14$ in Fig.~\ref{fig4}(e,f), respectively. A clear entanglement gap separating the low-lying PES levels from higher ones is identified. The number of PES levels below this gap exactly matches the typical counting of quasiparticle excitations resulting from the generalized Pauli principle of the non-Abelian Pfaffian or anti-Pfaffian state~\cite{regnault2011} (at most 2 particles in 4 consecutive orbitals). The particle-hole symmetry is explicitly broken by the non-uniform quantum geometries here, thus the particle-hole Pfaffian state~\cite{Son2015} is less likely to be a competing phase compared to Pfaffian and anti-Pfaffian states~\cite{Rezayi2017}. The concrete nature of the non-Abelian state need more detailed examination by wave function overlap. From the correspondence between the PES and the quasiparticle excitations, the numerical results further suggest the non-Abelian nature of this half-filled state. 

\emph{Discussions---}The method studied here can be further generalized to Chern bands with band inversion resembling higher LL. Concretely, an ideal local band inversion is introduced with a simple $\mathbf{k}\cdot\mathbf{p}$ model~\cite{tan2024} 
\begin{equation}
    \mathcal{H}_{\text{local}}(\mathbf{k}) =
    \begin{pmatrix}
    \alpha\mathbf{k}^2+\Delta & v_Fk_+ \\
    v_Fk_- & -\beta\mathbf{k}^2-\Delta
    \end{pmatrix},
\end{equation}
where $k_\pm=k_x\pm ik_y$. When $\Delta=-v_F^2/2(\alpha+\beta)$, both bands satisfy the trace and determinant conditions at any $\mathbf{k}$ with the quantum metric given by $g_{\mu\nu}(\mathbf{k})=(1/2)\left|\mathcal{B}(\mathbf{k})\right|\delta_{\mu\nu}$. Such an ideal local band inversion consistently yields a local Chern number $\mathcal{C}_{\text{local}}=1$ and a local average trace condition $\mathbb{T}_{\text{local}}=1$ when integrated over the region surrounding the inversion point. This framework explains why the orbitals $s$, $p_-$, $p_+$ form a complete basis for describing the LLL and 1LL, as illustrated in Fig.~\ref{fig1} and Fig.~\ref{fig3}. For the LCB, a single band inversion occurs between the $s$ and $p_-$ orbitals at the $\Gamma$ point. As a result, both $\mathcal{B}(\mathbf{k})$ and $\text{Tr}[g(\mathbf{k})]$ are concentrated around $\Gamma$, and the integral of them around $\Gamma$ yield a Chern number $\mathcal{C}=1$ and an average trace condition $\mathbb{T}\approx1$. For the 1CB, two band inversions occur: one between $p_-$ and $s$ at $\Gamma$, and the other between $p_-$ and $p_+$ at $K$. The integral of $\mathcal{B}(\mathbf{k})$ around $\Gamma$ gives $-1$, and around $K$, it gives $+1$; the corresponding local $\mathbb{T}$ are both close to $1$. This results in a net Chern number $\mathcal{C}=2-1=1$, and a total trace condition $\mathbb{T}=\left|2\right|+\left|-1\right|=3$. This construction can be extended to higher Chern bands. As shown in Fig.~\ref{fig1}(b,e), the band inversion at $M$ between $p_+$ and upper $d$ orbital contributes a local Chern number $\mathcal{C}_{\text{local}}=1$ and $\mathbb{T}_{\text{local}}=1$ at each $M$. Combined with the band inversion between $p_+$ and $p_-$ at $K$, this yields a total Chern number $\mathcal{C}=3-2=1$ and $\mathbb{T}=\left|3\right|+\left|-2\right|=5$ for the second Chern bands (i.e., the third band from the bottom), which matches the ideal value for the second LL. These results show that the quantum geometry of the constructed Chern bands closely matches that of generalized LL~\cite{liu2025}. It is worth mentioning that the constructed Chern bands only have a perfect quantum weight at the penalty of Berry curvature flatness, since $\mathcal{B}(\mathbf{k})$ always concentrates at the local band inversion point. To obtain the FCI state in the partially filled Chern bands within exact diagonalization, a delicate balance between $\mathcal{B}(\mathbf{k})$ flatness and ideal $\mathbb{T}$ should be considered. 

We also examine a minimal model that incorporates short-range hopping up to third-nearest neighbors, yielding a similar band structure and quantum geometry. Importantly, the many-body topological states remain robust in this setting~\cite{supple}, making it suitable to implementation in cold atom systems~\cite{leonard2023,Cooper2019,Aidelsburger2013,Miyake2013}, since only short-range density-density interactions are required. Optical lattices with synthetic gauge fields offer a controllable platform for realizing such orbital-based models with tunable interactions. Crucially, the lattice formulation is not a mere simplification but a central advantage, opening a practical route toward experimental realizations of fractional Chern insulators and non-Abelian phases in ultracold gases.

We note that the orbital construction of LL presented here is consistent with the concept of higher vortexability for the 1LL, as proposed in~\cite{fujimoto2024}. Specifically, the 1CB in our model is not self-vortexable, due to its large quantum metric trace $\mathbb{T}$. However, when the LCB and 1CB are considered together, we observe a nearly ideal combined trace $\mathbb{T}=2.2$ for the two lowest bands. This behavior originates from the band inversion between the $p_-$ and $p_+$ at the $K$ point. Our results thus offer a new perspective on understanding the vortexability of the lowest few LLs from a lattice orbital viewpoint.

Our orbital-based construction of LLs provides a complementary approach to QH lattice models. The well-known Kapit–Mueller (KM) model offers an exact lattice realization of LLL wavefunctions through engineered hopping terms~\cite{Kapit2010,Shen2025}. In contrast, our scheme relies on sequential band inversions among local Wannier orbitals of $s$, $p_-$, $p_+$ character. This mechanism not only reproduces the LLL but also generates higher LLs within a unified multiband framework, thereby offering direct access to non-Abelian states inherent to higher levels.

\begin{acknowledgments}
\emph{Acknowledgments---}We thank Biao Lian and Dung-Hai Lee for helpful discussions. This work is supported by the Natural Science Foundation of China through Grants No.~12350404 and No.~12174066, the Innovation Program for Quantum Science and Technology through Grant No.~2021ZD0302600, the Science and Technology Commission of Shanghai Municipality under Grants No.~23JC1400600, No.~24LZ1400100 and No.~2019SHZDZX01, and is sponsored by the ``Shuguang Program'' supported by the Shanghai Education Development Foundation and Shanghai Municipal Education Commission. 

H.W. and R.S. contributed equally to this work.
\end{acknowledgments}

\end{document}


\title{Supplementary Materials for ``Orbital Description of Landau Levels''}
\author{Huan Wang}
\affiliation{State Key Laboratory of Surface Physics and Department of Physics, Fudan University, Shanghai 200433, China}
\affiliation{Shanghai Research Center for Quantum Sciences, Shanghai 201315, China}
\author{Rui Shi}
\affiliation{State Key Laboratory of Surface Physics and Department of Physics, Fudan University, Shanghai 200433, China}
\affiliation{Shanghai Research Center for Quantum Sciences, Shanghai 201315, China}
\author{Zhaochen Liu}
\affiliation{State Key Laboratory of Surface Physics and Department of Physics, Fudan University, Shanghai 200433, China}
\affiliation{Shanghai Research Center for Quantum Sciences, Shanghai 201315, China}
\author{Jing Wang}
\thanks{Corresponding author: wjingphys@fudan.edu.cn}
\affiliation{State Key Laboratory of Surface Physics and Department of Physics, Fudan University, Shanghai 200433, China}
\affiliation{Shanghai Research Center for Quantum Sciences, Shanghai 201315, China}
\affiliation{Institute for Nanoelectronic Devices and Quantum Computing, Fudan University, Shanghai 200433, China}
\affiliation{Hefei National Laboratory, Hefei 230088, China}
\maketitle

\tableofcontents

\newpage

\section{Projected Three-Orbital Tight-Binding Model}
As seen in the main text, the band structure of continuum model in the dispersive Landau level limit ($\lambda=1$) does not possess a local gap below which the total Chern number is zero. This brings problems to the Wannier projection of LLL and 1LL. To avoid the Wannier obstruction, we first extract the three bands with $s$, $p_-$, $p_+$ orbital characteristics from the lowest few bands and then construct the standard maximally localized Wannier functions (MLWFs). The first step is done with the help of band disentanglement algorithm, where we use lowest four bands as input and freeze the lowest target two bands (LLL and 1LL). 

\subsection{Parameters for tight-binding model}
\begin{figure}[b]
	\begin{center}
		\includegraphics[width=0.7\columnwidth, clip=true]{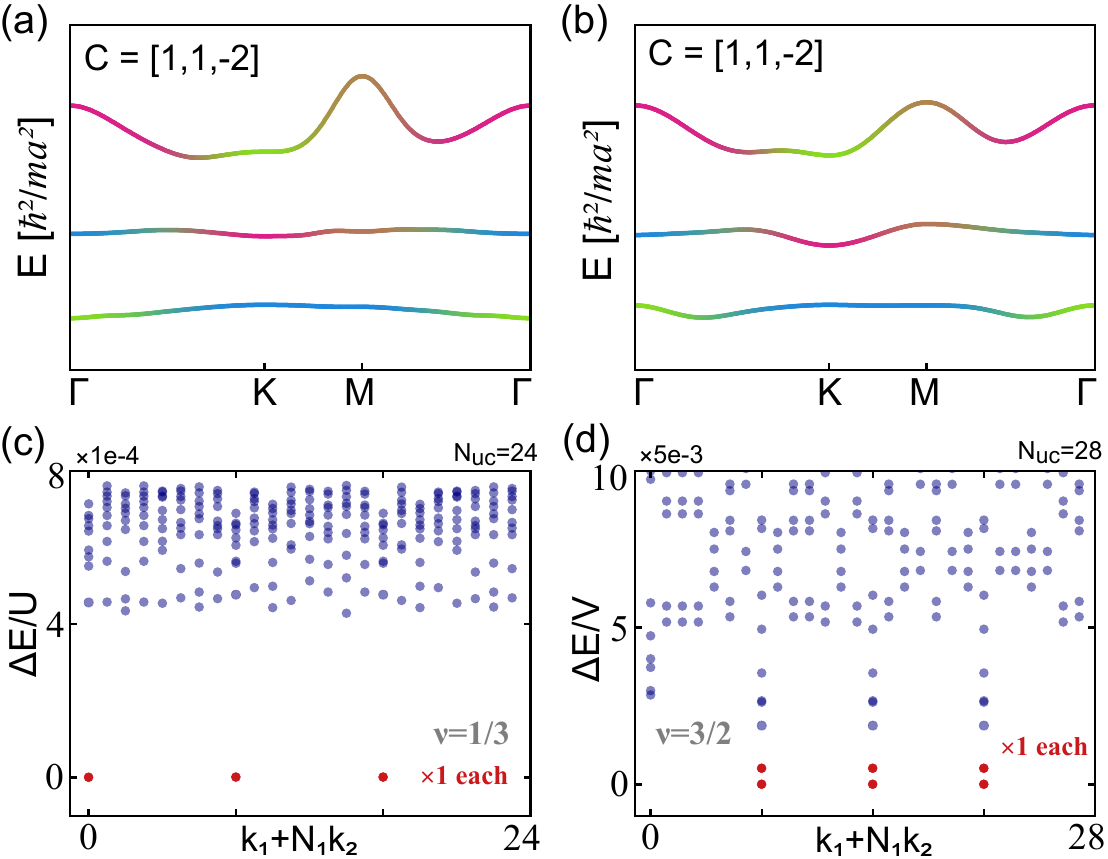}
	\end{center}
	\caption{ Wannier projected tight-binding model with (a) all long-range hopping terms; (b) Retain up to third nearest neighbor hopping terms. (c) Low energy many-body energy spectrum for $1/3$-filled LCB of reduced tight-binding model. (d) Energy spectrum for $1/2$-filled 1CB of reduced tight-binding model.}
	\label{figS1}
\end{figure}

The three-orbital tight-binding model is constructed from the continuum model by projecting the lowest three bands. And the resulting Hamiltonian including MLWFs with $s, p_{-}, p_{+}$ orbital charateristics can be expressed as
\begin{equation}
	\begin{aligned}
		\mathcal{H}_{\text{TB}} =& \sum_{i}\sum_{n=0}^{5} \left(t_{ss}^{(\alpha, \beta, n)}c^\dagger_{\mathbf{r}_i + C_6^n(\alpha,\beta),s}c_{\mathbf{r}_i,s} + t_{p_{-}p_{-}}^{(\alpha, \beta, n)}c^\dagger_{\mathbf{r}_i + C_6^n(\alpha,\beta),p_{-}}c_{\mathbf{r}_i,p_{-}} + t_{p_{+}p_{+}}^{(\alpha, \beta, n)}c^\dagger_{\mathbf{r}_i + C_6^n(\alpha,\beta),p_{+}}c_{\mathbf{r}_i,p_{+}}\right)\\
		& + \sum_{i}\sum_{n=0}^{5} \left(t_{sp_{-}}^{(\alpha, \beta, n)}c^\dagger_{\mathbf{r}_i + C_6^n(\alpha,\beta),s}c_{\mathbf{r}_i,p_{-}} + t_{sp_{+}}^{(\alpha, \beta, n)}c^\dagger_{\mathbf{r}_i + C_6^n(\alpha,\beta),s}c_{\mathbf{r}_i,p_{+}} + t_{p_{-}p_{+}}^{(\alpha, \beta, n)}c^\dagger_{\mathbf{r}_i + C_6^n(\alpha,\beta),p_{-}}c_{\mathbf{r}_i,p_{+}} + \text{h.c.}\right).
	\end{aligned}
\end{equation}
Here $c^\dagger_{\mathbf{r}_i, \gamma}/c_{\mathbf{r}_i, \gamma}$ creates/annihilates an electron with orbital $\gamma$ on site $\mathbf{r}_i$ and $\mathbf{r}_i + C_6^n(\alpha,\beta) \equiv \mathbf{r}_i + C_6^n(\alpha \mathbf{a}_1 + \beta \mathbf{a}_2)$ in which $C_6^n$ is six-fold rotational operator. $t_{\gamma, \gamma'}^{(\alpha, \beta, n)}$ is the hopping amplitude between orbital $\gamma$ on site $\mathbf{r}_i$ to orbital $\gamma'$ on site $\mathbf{r}_i + C_6^n(\alpha \mathbf{a}_1 + \beta \mathbf{a}_2)$. Due to the six-fold rotational symmetry, we have the following relations between hopping amplitude,
\begin{equation}
	\begin{aligned}
		t_{ss}^{(\alpha, \beta, n)} = t_{ss}^{(\alpha, \beta, 0)},\qquad t_{p_{-}p_{-}}^{(\alpha, \beta, n)} =& t_{p_{-}p_{-}}^{(\alpha, \beta, 0)},\qquad t_{p_{+}p_{+}}^{(\alpha, \beta, n)} = t_{p_{+}p_{+}}^{(\alpha, \beta, 0)}\\
		t_{sp_{-}}^{(\alpha, \beta, n)} = e^{-i \frac{\pi n}{3}}t_{sp_{-}}^{(\alpha, \beta, 0)},\qquad t_{sp_{+}}^{(\alpha, \beta, n)} =& e^{i \frac{\pi n}{3}}t_{sp_{+}}^{(\alpha, \beta, 0)},\qquad t_{p_{-}p_{+}}^{(\alpha, \beta, n)} = e^{i \frac{2\pi n}{3}}t_{p_{-}p_{+}}^{(\alpha, \beta, 0)}
	\end{aligned}
\end{equation}
which holds for $n=0, \cdots, 5$. The tight-binding model is truncated up to hopping range $5 \mathbf{a}_1$ and the parameters is given in Table~\ref{parameters}.

\begin{table}[h]
	\caption{Hopping strength parameters for Tight-Binding model.}
	\begin{center}\label{parameters}
		\renewcommand{\arraystretch}{1.5}
		\begin{tabular*}{\columnwidth}
			{@{\extracolsep{\fill}}ccccccc}
			\hline
			\hline
			$(\alpha, \beta)$ & $(0, 0)$ & $(1, 0)$ & $(1, 1)$ & $(2, 0)$ & $(2, 1)$ & $(1, 2)$ \\
			\hline
			$t_{ss}^{(\alpha, \beta, 0)}$ & $-39.176$ & $1.463$ & $0.068$ & $-0.112$ & $0.012$ & $0.012$  \\
			$t_{sp_{-}}^{(\alpha, \beta, 0)}$ & $0$ & $-2.278 + 0.859i$ & $0.101-0.124i$ & $0.151-0.057i$ & $0.006+0.070i$ & $-0.070+0.008i$  \\
			$t_{sp_{+}}^{(\alpha, \beta, 0)}$ & $0$ & $-0.285-0.473i$ & $-0.006-0.332i$ & $0.125+0.207i$ & $0.012-0.018i$ & $-0.012-0.017i$  \\
			$t_{p_{-}p_{-}}^{(\alpha, \beta, 0)}$ & $-37.254$  & $-3.540$ & $0.354$ & $0.640$ & $-0.174$ & $-0.174$  \\
			$t_{p_{-}p_{+}}^{(\alpha, \beta, 0)}$ & $0$  & $-0.487-2.654i$ & $0.044-0.037i$ & $0.052+0.286i$ & $0.164+0.107i$ & $-0.079-0.179i$\\
			$t_{p_{+}p_{+}}^{(\alpha, \beta, 0)}$ & $-36.024$  & $2.030$ & $-0.683$ & $0.677$ & $-0.062$ & $-0.062$ \\
			\hline
			\hline
			$(\alpha, \beta)$ & $(3, 0)$ & $(3, 1)$ & $(1, 3)$ & $(2, 2)$ & $(4, 0)$ & $(4, 1)$ \\
			\hline
			$t_{ss}^{(\alpha, \beta, 0)}$ & $0.022$ & $-0.003$  & $-0.003$  & $-0.014$ & $-0.010$ & $0.002$  \\
			$t_{sp_{-}}^{(\alpha, \beta, 0)}$ & $-0.065+0.025i$ & $0.011-0.027i$ & $0.024-0.016i$ & $0.017-0.021i$ & $0.011-0.004i$ & $0.001+0.007i$ \\
			$t_{sp_{+}}^{(\alpha, \beta, 0)}$ & $-0.025-0.041i$ & $0.003-0.015i$ & $-0.003-0.015i$  & $0.001+0.050i$ & $0.012+0.020i$ & $-0.003-0.002i$\\
			$t_{p_{-}p_{-}}^{(\alpha, \beta, 0)}$ & $-0.164$ & $0.042$ & $0.042$  & $0.121$ & $0.042$ & $-0.005$\\
			$t_{p_{-}p_{+}}^{(\alpha, \beta, 0)}$ & $-0.040-0.218i$ & $0.008+0.023i$ & $-0.021-0.012i$ & $-0.044+0.038i$ & $0.006+0.030i$ & $0.008+0.006i$\\
			$t_{p_{+}p_{+}}^{(\alpha, \beta, 0)}$ & $0.118$ & $-0.078$ & $-0.078$  & $0.084$ & $0.061$ & $-0.013$\\
			\hline
			\hline
			$(\alpha, \beta)$ & $(3, 2)$ & $(2, 3)$ & $(1, 4)$ & $(5, 0)$ & \\
			\hline
			$t_{ss}^{(\alpha, \beta, 0)}$ & $0.005$ & $0.005$ & $0.002$ & $0.003$ & \\
			$t_{sp_{-}}^{(\alpha, \beta, 0)}$ & $-0.014+0.007i$ & $-0.004+0.015i$ & $-0.007$ & $-0.006+0.002i$ & \\
			$t_{sp_{+}}^{(\alpha, \beta, 0)}$ & $-0.002-0.006i$ & $0.002-0.006i$ & $0.003-0.002i$ & $-0.002-0.003i$ & \\
			$t_{p_{-}p_{-}}^{(\alpha, \beta, 0)}$ & $-0.037$ & $-0.037$  & $-0.005$ & $-0.021$ & \\
			$t_{p_{-}p_{+}}^{(\alpha, \beta, 0)}$ & $-0.013-0.031i$ & $0.029+0.018i$  & $-0.004-0.009i$ & $-0.003-0.018i$ & \\
			$t_{p_{+}p_{+}}^{(\alpha, \beta, 0)}$  & $0.014$ & $0.014$ & $-0.013$ & $0.009$ & \\
			\hline
			\hline
		\end{tabular*}
	\end{center}
\end{table}


The band structure with all long-range hopping terms is shown in Fig.~\ref{figS1}(a), while the band structure with up to third nearest neighbor hopping terms is shown in Fig.~\ref{figS1}(b). We find that the Berry curvature fluctuation $\delta\mathbf{\mathcal{B}}$ and average trace condition $\mathbb{T}$ for LCB and 1CB in Fig.~\ref{figS1}(b) are $\delta\mathcal{B}_{\text{LCB}}=0.77$, $\mathbb{T}_{\text{LCB}}=1.562$ and $\delta\mathcal{B}_{\text{1CB}}=1.64$, $\mathbb{T}_{\text{1CB}}=2.849$, respectively. Remarkably, they are very close to the those values in original model containing all long-range hopping terms (see Table~\ref{Topo_Compare_tab}). However, these values will significantly deviate from above values if we only keep up to next-nearest-neighbor hopping terms. Moreover, we have done the exact diagonalization in this model with only third nearest neighbor hopping terms, and find $1/3$-filled LCB supports Abelian state (see Fig.~\ref{figS1}(c)), while half-filled 1CB supports non-Abelian state (see Fig.~\ref{figS1}(d)). 

\renewcommand{\arraystretch}{1.5}
\begin{table}[!h]
        \centering
        
\begin{tabular}{p{0.06\textwidth}p{0.06\textwidth}p{0.06\textwidth}p{0.06\textwidth}|p{0.06\textwidth}p{0.06\textwidth}p{0.06\textwidth}p{0.06\textwidth}}
\hline 
\hline 
 \multicolumn{4}{c|}{
$\mathcal{H}_{\text{TB}}$ with long range hopping
} & \multicolumn{4}{c}{
$\mathcal{H}_{\text{TB}}$ with short range hopping
} \\
\hline 
 \multicolumn{2}{c}{
LCB
} & \multicolumn{2}{c|}{
1CB
} & \multicolumn{2}{c}{
LCB
} & \multicolumn{2}{c}{
1CB
} \\
\hline 
\centering $\delta\mathcal{B}$  & \centering $\mathbb{T}$ & \centering $\delta\mathcal{B}$ & \centering $\mathbb{T}$ & \centering $\delta\mathcal{B}$ & \centering $\mathbb{T}$ & \centering $\delta\mathcal{B}$ & \hspace{1.5em}$\mathbb{T}$ \\
\hline
\centering 0.78  & \centering 1.543 & \centering 1.51 & \centering 2.843 & \centering 0.77 & \centering 1.562 & \centering 1.64 & \hspace{0.7em}2.849 \\
\hline
\hline
\end{tabular}
\caption{Berry curvature fluctuation $\delta\mathcal{B}$
and average trace condition $\mathbb{T}$ for LCB and 1CB in $\mathcal{H}_{\text{TB}}$ with different hopping range. Band structure of $\mathcal{H}_{\text{TB}}$ with long range hopping (up to 15th nearest neighbor) is shown in Fig.~\ref{figS1}(a), Band structure of $\mathcal{H}_{\text{TB}}$ with short range hopping (up to 3rd nearest neighbor) is shown in Fig.~\ref{figS1}(b).}
\label{Topo_Compare_tab}
\end{table}
\renewcommand{\arraystretch}{1}

\section{Projected Interaction}
\subsection{Parameters of onsite and nearest-neighbor interaction}
After we got the projected tight-binding model, we need to include the interaction to study many body physics at fractional filling in LCB and 1CB. Since we have the Wannier functions $w_s$, $w_{p_-}$, $w_{p_+}$ with $s$, $p_-$, $p_+$ orbital characteristics, the natural way to estimate the interaction is to insert the expansion
\begin{equation}
\Psi^\dagger(\mbf{r})=w^*_s(\mbf{r}) c^\dagger_s + w^*_{p_-}(\mbf{r}) c^\dagger_{p_-} + w^*_{p_+}(\mbf{r}) c^\dagger_{p_+},
\end{equation}
into the general expression of the Coulomb interaction
\begin{equation}
  \mathcal{H}_{\text{Coulomb}}=\frac{1}{2}\int d\mbf{r}\int d\mbf{r}^\prime \Psi^\dagger(\mbf{r}) \Psi^\dagger(\mbf{r}^\prime) \frac{e^2}{\left|\mbf{r}-\mbf{r}^\prime\right|} \Psi(\mbf{r}^\prime) \Psi(\mbf{r}),
\end{equation}
we could expect 81 terms in general. First we have from the choice \emph{abba} (i.e., choosing the \emph{a}-term from $\Psi^\dagger(\mbf{r})$, the \emph{b}-term from $\Psi^\dagger(\mbf{r}^\prime)$ and $\Psi(\mbf{r}^\prime)$, and again the \emph{a}-term from $\Psi(\mbf{r})$), here \emph{a} and \emph{b} are spanned by $s$, $p_-$, $p_+$ orbital index.
\begin{equation}
  \frac{1}{2}\int d\mbf{r}\int d\mbf{r}^\prime w^*_a(\mbf{r}) w^*_b(\mbf{r}^\prime) \frac{e^2}{\left|\mbf{r}-\mbf{r}^\prime\right|} w_b(\mbf{r}^\prime) w_a(\mbf{r}) c^\dagger_ac^\dagger_bc_bc_a \rightarrow \frac{C_{ab}}{2} n_an_b,
\end{equation}
\emph{baab} gives the same form as \emph{abba}. Direct exchange contributions arise from \emph{abab} and \emph{baba}, e.g.,
\begin{equation}
  \frac{1}{2}\int d\mbf{r}\int d\mbf{r}^\prime w^*_a(\mbf{r}) w^*_b(\mbf{r}^\prime) \frac{e^2}{\left|\mbf{r}-\mbf{r}^\prime\right|} w_a(\mbf{r}^\prime) w_b(\mbf{r}) c^\dagger_ac^\dagger_bc_ac_b \rightarrow -\frac{J_{ab}}{2} n_an_b,
\end{equation}
So the onsite interaction $U_{ab}$ in the maintext can be estimated as $C_{ab}-J_{ab}$, the other terms in onsite interaction are found to zero because of the existence of $C_6$ symmetry.

\renewcommand{\arraystretch}{1.4}
\begin{table}[t]
  \begin{center}   
    \caption{Estimated onsite interaction $U_{ab}=C_{ab}-J_{ab}$ from Wannier functions, here the values are in the unit of $(e^2/\eps_0 a_0)$.}
    \label{tab1} 
    \begin{tabular}{cm{2.0cm}<{\centering}m{2.0cm}<{\centering}m{2.0cm}<{\centering}} 
  \hline
  \hline    & $n_sn_{p_-}$ & $n_sn_{p_+}$ & $n_{p_-}n_{p_+}$ \\   
  \hline  $U_{ab}$ & 1.71 & 1.52  & 1.57 \\  
  \hline  
  \hline 
    \end{tabular}   
  \end{center}   
\end{table}
\renewcommand{\arraystretch}{1}

\renewcommand{\arraystretch}{2}
\begin{table}[b]  
  \begin{center}   
    \caption{The absolute value of the estimated nearest-neighbor interactions $D_{abcd}$ from Wannier functions, here these values are in the unit of $(e^2/\eps_0 a_0)$, and the dominant density-density terms are labeled as blue.}
    \label{tab2} 
    \begin{tabular}
			{m{1.2cm}<{\centering}m{1.2cm}<{\centering}m{1.2cm}<{\centering}m{1.2cm}<{\centering}m{1.2cm}<{\centering}m{1.2cm}<{\centering}m{1.2cm}<{\centering}m{1.2cm}<{\centering}m{1.2cm}<{\centering}m{1.2cm}<{\centering}} 
  \hline      
  \hline $\left|D_{abcd}\right|$         & $c_sc_s$ & $c_sc_{p_-}$ & $c_sc_{p_+}$ & $c_{p_-}c_s$ & $c_{p_-}c_{p_-}$ & $c_{p_-}c_{p_+}$ & $c_{p_+}c_s$ & $c_{p_+}c_{p_-}$ & $c_{p_+}c_{p_+}$ \\   
  \hline $c^\dagger_sc^\dagger_s$        &   \textcolor{blue}{0.99}   &     0.19     &     0.19     &     0.19     &       0.09       &       0.04       &     0.19     &       0.04       &       0.10  \\  
  \hline $c^\dagger_sc^\dagger_{p_-}$    &   0.19   &     0.04     &     0.10     &     \textcolor{blue}{1.00}     &       0.19       &       0.19       &     0.09     &       0.02       &       0.06  \\
  \hline $c^\dagger_sc^\dagger_{p_+}$    &   0.19   &     0.10     &     0.03     &     0.09     &       0.06       &       0.01       &     \textcolor{blue}{1.03}     &       0.19       &       0.20  \\
  \hline $c^\dagger_{p_-}c^\dagger_s$    &   0.19   &     \textcolor{blue}{1.00}     &     0.09     &     0.04     &       0.18       &       0.02       &     0.10     &       0.19       &       0.06  \\
  \hline $c^\dagger_{p_-}c^\dagger_{p_-}$&   0.09   &     0.18     &     0.06     &     0.19     &       \textcolor{blue}{0.99}       &       0.09       &     0.06     &       0.09       &       0.05  \\  
  \hline $c^\dagger_{p_-}c^\dagger_{p_+}$&   0.04   &     0.19     &     0.01     &     0.02     &       0.09       &       0.01       &     0.19     &       \textcolor{blue}{1.03}       &       0.08  \\
  \hline $c^\dagger_{p_+}c^\dagger_s$    &   0.19   &     0.09     &     \textcolor{blue}{1.03}     &     0.10     &       0.06       &       0.19       &     0.03     &       0.01       &       0.20  \\
  \hline $c^\dagger_{p_+}c^\dagger_{p_-}$&   0.04   &     0.02     &     0.19     &     0.19     &       0.09       &       \textcolor{blue}{1.03}       &     0.01     &       0.01       &       0.08  \\
  \hline $c^\dagger_{p_+}c^\dagger_{p_+}$&   0.10   &     0.06     &     0.20     &     0.06     &       0.05       &       0.08       &     0.20     &       0.08       &       \textcolor{blue}{1.06}  \\
  \hline  
  \hline  
    \end{tabular}   
  \end{center}   
\end{table}
\renewcommand{\arraystretch}{1}

However, when it come to nearest-neighbor interaction, all of the 81 terms can be nonzero, the general choice \emph{abcd} is 
\begin{equation}
  \frac{1}{2}\int d\mbf{r}\int d\mbf{r}^\prime w^*_a(\mbf{r}) w^*_b(\mbf{r}^\prime) \frac{e^2}{\left|\mbf{r}-\mbf{r}^\prime\right|} w_c(\mbf{r}^\prime) w_d(\mbf{r}) c^\dagger_ac^\dagger_bc_cc_d \rightarrow \frac{D_{abcd}}{2} c^\dagger_ac^\dagger_bc_cc_d,
\end{equation}
In Table~\ref{tab1} and Table~\ref{tab2} we list the estimated onsite interaction $U_{ab}=C_{ab}-J_{ab}$ and absolute value of nearest-neighbor interaction $D_{abcd}$, respectively. We can see that the density-density interactions are always the dominant terms, so we only keep the leading terms, namely the onsite and nearest-neighbor density-density interactions.

\begin{figure}[h]
	\begin{center}
		\includegraphics[width=0.7\columnwidth, clip=true]{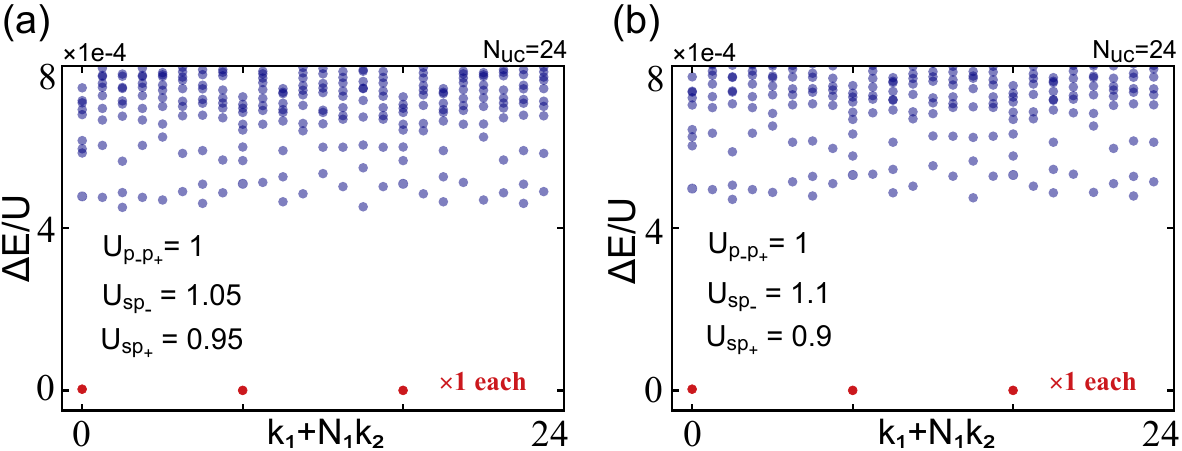}
	\end{center}
	\caption{Low-energy many-body energy spectrum for $1/3$-filled LCB. (a) $5\%$ anisotropy (b) $10\%$ anisotropy in the density-density interactions for different orbital pairs.}
	\label{figS2}
\end{figure}

Although the dominant density-density interactions are different for different orbital pairs, their values are only slightly deviating from the average value. In the main text, we adopted an isotropic approximation, neglecting minor variations in the interactions between different orbital pairs. To justify this assumption, we present the many-body spectra of the $1/3$-filled LCB with $5\%$ and $10\%$ anisotropy of the density-density interaction in Fig.~\ref{figS2}, respectively. We find that both spectra exhibit the same ground state as the isotropic case shown in Fig.~4(a) of the main text. Therefore, for simplicity, we utilize the isotropic density interaction for the minimal model. Moreover, for the LCB, we set $U=1$, and ignore the nearest-neighbor term. While for the 1CB, the long-range interaction is crucial for the non-Abelian state. To account for the long-range effects that were initially underestimated with nearest-neighbor cut off, we set $U/V=1.3$.


\section{Characterization of the $\nu=1/3$ and $\nu=3/2$ States}
\subsection{Exact diagonalization and particle entanglement spectrum}
Exact Diagonalization (ED) is a useful numerical tool to detect the ground state degeneracy which is one of prominent features of fractional Chern insulator. Particle entanglement spectrum (PES) which encodes the information of the quasihole excitations is quite different between topological and trivial states.

\subsubsection{$\nu=1/3$}
\begin{figure}[hbtp]
	\begin{center}
    \includegraphics[width=0.98\columnwidth, clip=true]{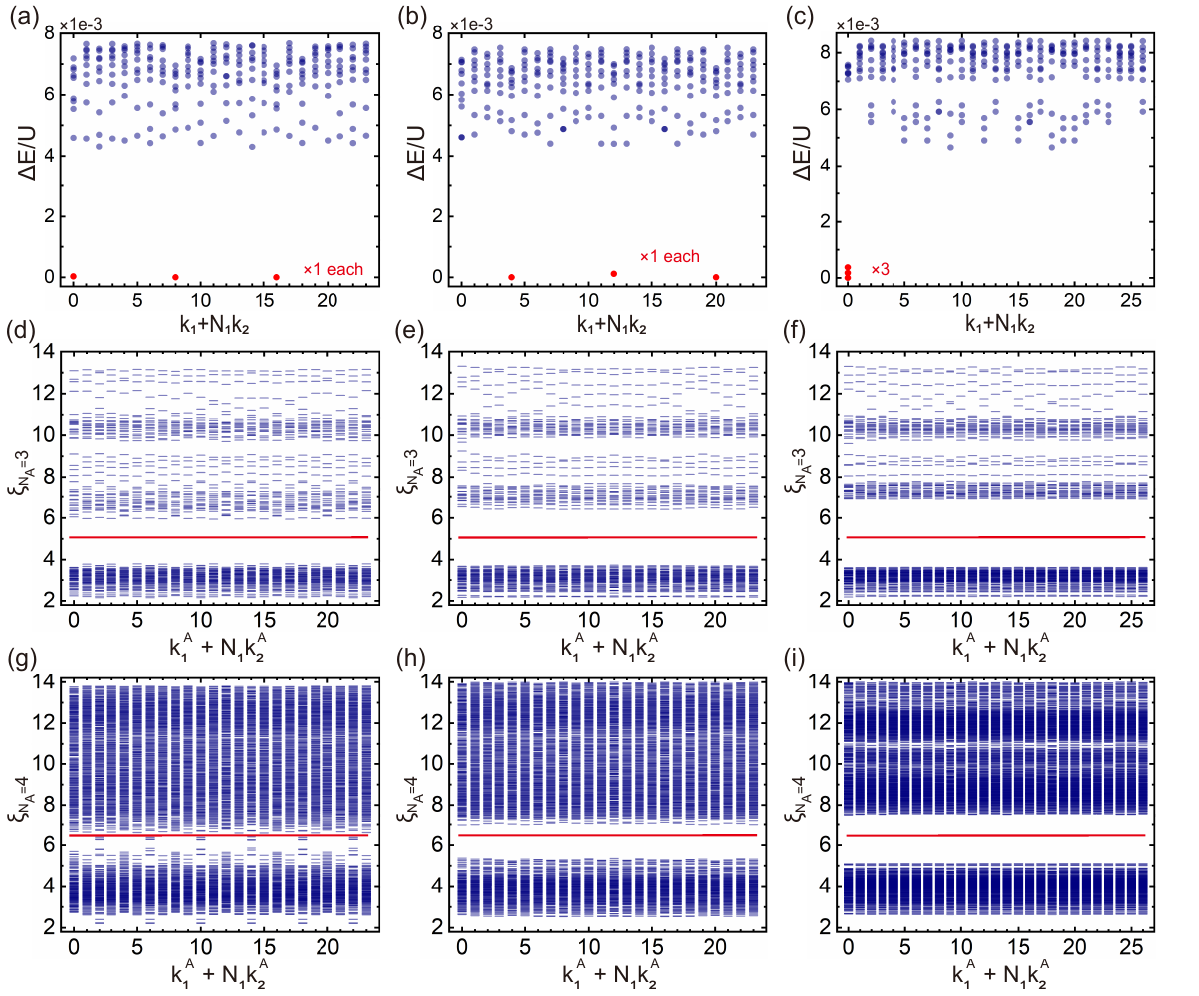}
	\end{center}
	\caption{Many-body spectra and particle entanglement spectra of fractional filled LCB ($\nu = 1/3$) for different cluster geometries. (a),(d),(g) system size $N_{\text{uc}}=24(A)$; (b),(e),(h) system size $N_{\text{uc}}=24(B)$; (c),(f),(i) system size $N_{\text{uc}}=27$. The number of levels under the red line in PES has the right counting originating from generalized Pauli principle.}
  \label{figS3}
\end{figure}

 In main text, we show the many-body spectrum for filling $\nu = 1/3$ in LCB on system size $N_{\text{uc}}=24(A)$ (see detail in Sec.~\ref{Cluster_Geometry}). To exclude the finite size effect, we further show the low-energy spectra at filling $\nu = 1/3$ on system size $N_{\text{uc}}=24(B)$ and $N_{\text{uc}}=27$ in Fig.~\ref{figS3}. The results are plotted as a function of crystal momentum $\mathbf{k}=k_1\mathbf{T}_1 + k_1\mathbf{T}_2$ which is labeled as $k=k_1 + N_1k_2$ where $k_{1(2)}=0, \cdots, N_{1(2)}-1$ for system size $N_{\text{uc}}=N_1\times N_2$ with filled particle number $N_e=\nu N_{\text{uc}}$ and $\mathbf{T}_i$ are basis vectors of crystal momentum defined in Sec.~\ref{Cluster_Geometry}. For these three sizes, there exhibits three-fold nearly degenerate ground states which are well separated from excited states. A clear gap remains in different cluster geometry indicating its existence in thermodynamic limit. Significantly, The lattice momenta of the degenerate ground states have linear indices $(0, 8, 16)$, $(4, 12, 20)$, $(0, 0, 0)$ respectively and are all in precise agreement with the generalized Pauli principle which is the hallmark of FCI at $1/3$ filling.

In order to distinguish FCI and other competing phases, and further confirm the existence of FCI, the PES is calculated. Specifically, we divide the whole system into two subsystems of $N_A$ and $N_B$ particles, then trace out part $B$. The PES levels $\xi_i$ is associated with the eigenvalues $\lambda_i$ of reduced density matrix $\rho_A=$Tr${}_B \rho$ through $\xi_i =- $log$\lambda_i$ where $\rho$ is defined as $(1/d)\sum_{i}^{d}|\Psi_i\rangle \langle \Psi_i|$ in which $|\Psi_i\rangle$ are degenerate many-body ground states and $d$ is ground state degeneracy. For $N_A=3$, the number of levels below the entanglement gap is $1088, 1088, 1710$ for size $N_{\text{uc}}=24(A), 24(B), 27$ respectively. And for $N_A=4$, the number of levels below the entanglement gap is $2730, 2730, 5508$ for size $N_{\text{uc}}=24(A), 24(B), 27$ respectively. All the counting results exactly matches the counting rule of $1/3$ Laughlin state resulting from generalized Pauli principle. Furthermore, We find the entanglement gap separating the low-lying PES levels from higher ones is augmented with more uniform cluster geometry and the larger system size indicating the stability of $1/3$ FCI state in thermodynamic limit.

\subsubsection{$\nu=3/2$}
\begin{figure}[htb]
	\begin{center}
		\includegraphics[width=0.92\columnwidth, clip=true]{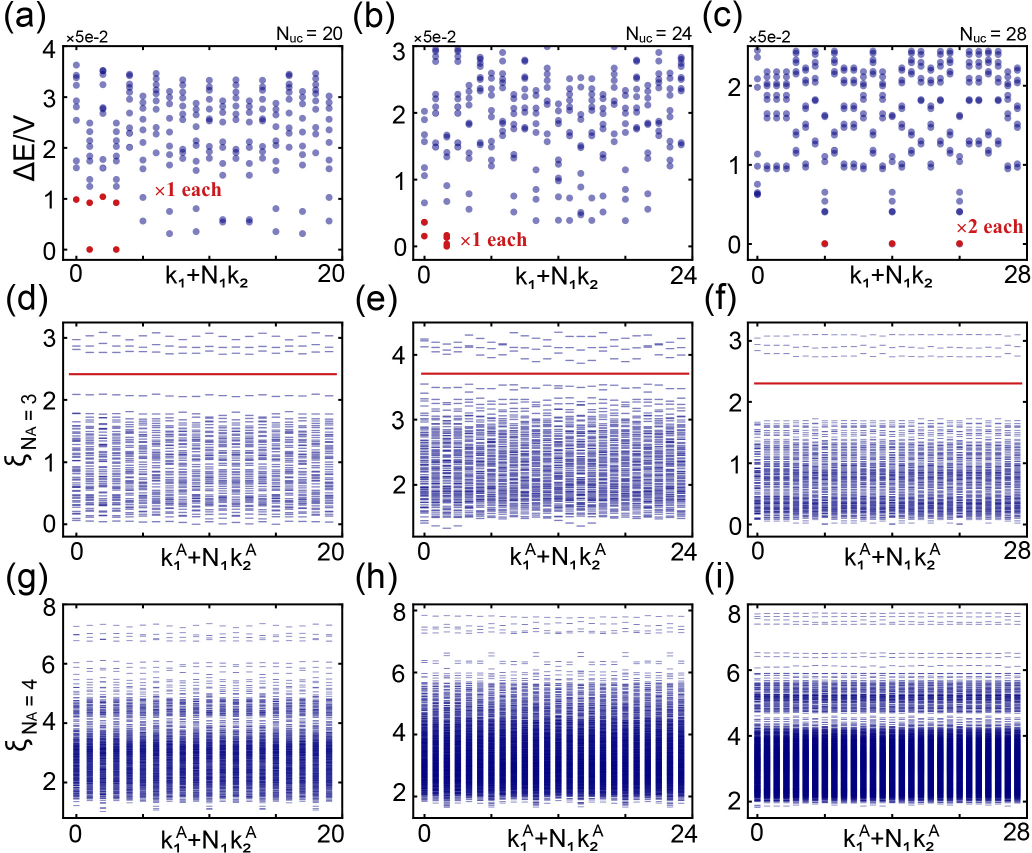}
	\end{center}
	\caption{Many-body spectra and particle entanglement spectra of half-filled 1CB ($\nu = 3/2$) for different cluster geometries. (a),(d),(g) system size $N_{\text{uc}}=20$; (b),(e),(h) system size $N_{\text{uc}}=24(A)$; (c),(f),(i) system size $N_{\text{uc}}=28$. The number of levels under the red line in entanglement spectra has the right counting originating from generalized Pauli principle.}
	\label{figS4}
\end{figure}

In main text, we show the many-body spectrum for half-filled 1CB on system size $N_{\text{uc}}=26$ and $N_{\text{uc}}=28$. Here we further show the low-energy spectra at filling $\nu = 3/2$ on system size $N_{\text{uc}}=20$ and $N_{\text{uc}}=24(A)$ in Fig.~\ref{figS4}. The results are plotted as a function of crystal momentum $k=k_1 + N_1k_2$. For size $N_{\text{uc}}=20$ and $N_{\text{uc}}=24(A)$, the possible FCI state is labeled as red according to generalized Pauli principle. We can see as the size increases, the gap between FCI state and other states appear in size $N_{\text{uc}}=28$, which indicating the stability of $3/2$ FCI state in thermodynamic limit.

In order to distinguish FCI and other competing phases, and further confirm the existence of FCI, we calculated the PES for each size. For $N_A=3$, the number of levels below the entanglement gap is $1080, 1952, 3192$ for size $N_{\text{uc}}=20, 24(A), 28$ respectively which exactly matches the counting rule of Moore-Read state resulting from generalized Pauli principle. But for $N_A=4$, the entanglement gap is unclear and it's hard to count the number of levels below it. The gap in PES shrinks for the cases of $N_A>3$, which may indicate there are contributions from other low-energy competing states.

\subsection{Twist boundary condition}

The accuracy of the ED method is inherently constrained by computational power. For example, the Hilbert space for a system with $N_{\text{uc}}=32$ has a dimension of approximately $19$ million, which is close to the upper limit of what the current ED methodology can handle. To assess the finite-size effect, an effective approach is to introduce twist boundary conditions as a perturbation to the system. Under these conditions, the momentum of a single particle is modified to $k_{i}\rightarrow k_i+(\theta_i/2\pi)\mathbf{T}_i$ which allows for a near-continuous scanning of all momenta within the Brillouin zone.
 
\begin{figure}[htb]
	\begin{center}
		\includegraphics[width=0.6\columnwidth, clip=true]{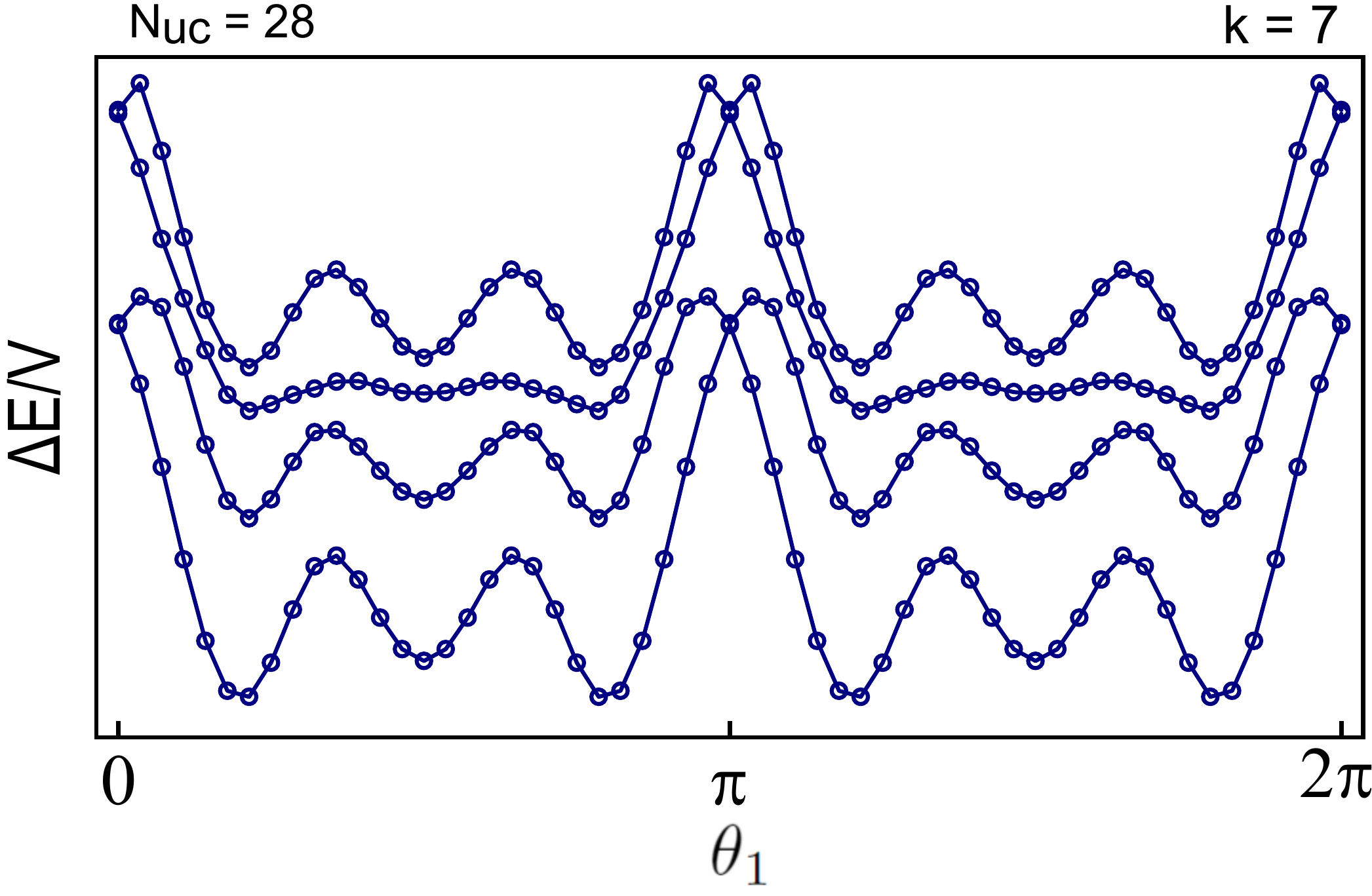}
	\end{center}
	\caption{Low-energy spectra at the linear indice $k$=(7) under the variation of the twist boundary phase
  $\theta_1$ for tight-binding model with system size $N_{\text{uc}}=28$.}
	\label{figS5}
\end{figure}

As shown in Fig.~\ref{figS5}, when tuning the boundary phases for a system size $N_{\text{uc}}=28$, the lowest two-fold nearly degenerated ground states in linear indice (7) sectors are generally separated from the excited states by direct gaps for given $k$ and boundary phase. The ground state energies remain well separated from the other low-energy excitation spectrum, indicating the robustness of the excitation gap.

\subsection{Lanlau level wave functions overlap}

As mentioned in main text, the wave functions overlap between the lowest two bands of the dispersive LLs and the corresponding flat LLs exceed 99\%. Here we briefly introduce the calculation method~\cite{wang2021,reddy2024}.

In order to compute the wave function overlap between the continuum model and Landau levels, we need to define the LL magnetic Bloch basis in torus by making use of magnetic translation algebra. The Hamiltonian of a particle moving in a uniform magnetic field $\mathbf{B}(\mathbf{r}) = -B_0\hat{z} = \nabla \times \mathbf{A}(\mathbf{r})$ can be written as
\begin{equation}
	\hat{H}_0 = \frac{(\mathbf{p} - e\mathbf{A})^2}{2m} = \omega_c (a^\dagger a + 1)
\end{equation}
where $a = \frac{l}{\sqrt{2}}(\bm{\pi}_x + i\bm{\pi}_y)$ and $\bm{\pi} = \bm{\pi} - e\mathbf{A}$ is canonical momentum, magnetic length $l = \sqrt{1/eB_0}$, cyclotron frequency $\omega_c = eB_0/m$. (Here and follows $\hbar = c = 1$) However, the Hamiltonian is not translation invariant under conventional translation operator $T(\mathbf{a})=e^{i\mathbf{a}\cdot \mathbf{p}}$ since the vector potential is not translation invariant in general. On the other hand, we can define the so-called magnetic translation operator~\cite{zak1966,girvin2019}
\begin{equation}
	t(\mathbf{a}) = e^{i\phi_a (\mathbf{r})} T(\mathbf{a})
\end{equation}
where $\phi_a$ is defined by $\mathbf{A}(\mathbf{r} + \mathbf{a}) - \mathbf{A}(\mathbf{r}) = \nabla \phi_a(\mathbf{r})$. Therefore, the Hamiltonian is translational invariant under magnetic translation operator, namely $[\hat{H}_0, t(\mathbf{a})]=0$. It's straightforward to verify the magnetic translation algebra
\begin{equation}
	t(\mathbf{a})t(\mathbf{a}') = e^{i\frac{1}{2l^2}\mathbf{a}\wedge \mathbf{a}'}t(\mathbf{a} + \mathbf{a}')
\end{equation}
where $\mathbf{a}\wedge \mathbf{a}' = (\mathbf{a}\times \mathbf{a}')\cdot \hat{z}$. In magnetic field, it is helpful to separate the motion of electron into guiding center $\mathbf{R}$ and cyclotron motion $\bar{\mathbf{R}}$
\begin{equation}
	\mathbf{R}^a = \mathbf{r}^a + \epsilon^{ab}\bm{\pi}_b l^2,\qquad \bar{\mathbf{R}}^a = -\epsilon^{ab}\bm{\pi}_b l^2
\end{equation}
with the commutation relation
\begin{equation}
	[\mathbf{R}^a, \mathbf{R}^b] = -i\epsilon^{ab}l^2,\qquad [\bar{\mathbf{R}}^a, \bar{\mathbf{R}}^b] = i\epsilon^{ab}l^2,\qquad [\mathbf{R}^a, \bar{\mathbf{R}}^b] = 0.
\end{equation}
If we choose a gauge where $\mathbf{A}(\mathbf{r})$ is linear in $\mathbf{r}$, we can rewrite magnetic translation operator as 
\begin{equation}
	t(\mathbf{a}) = e^{i\mathbf{a}\cdot \mathbf{P}}
\end{equation}
where $\mathbf{P} = \frac{1}{l^2}\hat{z}\times \mathbf{R}$. Then we can define the magnetic momentum boost operator
\begin{equation}
	\tau(\mathbf{q}) = e^{i\mathbf{q}\cdot \mathbf{R}},
\end{equation}
and the magnetic translation algebra can be expressed by $\tau(\mathbf{q})$ as
\begin{equation}
	\tau(\mathbf{q})\tau(\mathbf{q}') = e^{i\frac{l^2}{2}\mathbf{q}\wedge \mathbf{q}'}\tau(\mathbf{q} + \mathbf{q}').
\end{equation}
The relation between $t(\mathbf{a})$ and $\tau(\mathbf{q})$ is
\begin{equation}
	t(\mathbf{a}) = \tau(\mathbf{a}\times \hat{z}/l^2),\qquad t(\mathbf{a})\tau(\mathbf{q}) = e^{i\mathbf{q}\cdot \mathbf{a}}\tau(\mathbf{q})t(\mathbf{a})
\end{equation}

Now we define the primitive magnetic lattice vector $\mathbf{a}_i$ so that $\mathbf{a}_1\wedge\mathbf{a}_2 = 2\pi l^2$ (Note that $\mathbf{a}_i$ can be chosen arbitrarily as long as they enclose an area whose magnetic flux is a flux quantum). It is easy to prove that the set of operators $t(\mathbf{A})$ ($\mathbf{A} = m \mathbf{a}_1 + n \mathbf{a}_2 $ with $m,n$ integers) mutually commute. And $t(\mathbf{A})$ commute with Hamiltonian $\hat{H}_0$. So we can take advantage of magnetic Bloch theorem (analogy of Bloch theorem) to define the LL magnetic Bloch basis $|n, \mathbf{k}\rangle$. Starting with LLL magnetic Bloch state $|0, \mathbf{0}\rangle$, the general Bloch state is 
\begin{equation}
	|n, \mathbf{k}\rangle = |n\rangle\otimes |\mathbf{k}\rangle = \frac{(a^\dagger)^n}{\sqrt{n!}}|0\rangle \otimes \tau(\mathbf{k})|\mathbf{0}\rangle
\end{equation}
And the magnetic boundary condition is defined by $t(\mathbf{a}_i)|0, \mathbf{0}\rangle = e^{i2\pi\phi_i}|0, \mathbf{0}\rangle$. Next, we define the primitive magnetic reciprocal lattice vectors $\mathbf{b}_i = \epsilon_{ij}\mathbf{a}_j\times \hat{z}/l^2$ that satisfy $\mathbf{a}_i\cdot\mathbf{b}_j = 2\pi\delta_{ij}$ through $t(\mathbf{a}_i) = \tau(-\epsilon_{ij}\mathbf{b}_j)$. Since $|\mathbf{k}\rangle$ and $|\mathbf{k}+\mathbf{g}\rangle$ with $\mathbf{g} =g_1 \mathbf{b}_1 + g_2\mathbf{b}_2$ ($g_1, g_2$ are integers) satisfy the same boundary condition so they are the same state up to an overall phase. Therefore, the number of different states is equal to the number of allowed momentum module magnetic reciprocal lattice vector. The number is just the number of flux quantum threading the torus.

As far as the numerical calculation, we make the best of matrix element
\begin{equation}\label{eq14}
	\begin{aligned}
		&\langle m,\mathbf{p}|e^{i\mathbf{q}\cdot \mathbf{r}}|n,\mathbf{k}\rangle = \langle m|e^{i\mathbf{q}\cdot\bar{\mathbf{R}}}|n\rangle
		\langle \mathbf{p}|\tau(\mathbf{q})|\mathbf{k}\rangle\\
		=& e^{-l^2|\mathbf{q}|^2/4} G_{mn}(q) e^{i\frac{l^2}{2}(-\mathbf{p}\wedge(\mathbf{k+q}) + \mathbf{q}\wedge\mathbf{k})}\sum_\mathbf{g}\delta_{\mathbf{k+q-p}, \mathbf{g}}\eta(\mathbf{g})
	\end{aligned}
\end{equation}
where $\eta(\mathbf{g}) = \langle \mathbf{0}|\tau(\mathbf{g})|\mathbf{0}\rangle$ and 
\begin{equation}
	G_{mn}(q) = \begin{cases}
		q^{m-n}\sqrt{\frac{n!}{m!}}L_n^{m-n}(|q|^2),\qquad n\leq m\\
		(-\bar{q})^{n-m}\sqrt{\frac{m!}{n!}}L_m^{n-m}(|q|^2),\qquad n>m
	\end{cases}
\end{equation}
with $q = \frac{l(q_x + iq_y)}{\sqrt{2}}$ and $L_n^m(x)$ is associated Laguerre polynomials. The above formula can be proved by magnetic translation algebra. In calculation, we fix $\phi_1=\phi_2=1/2$ so that $\eta(\mathbf{g}) = 1$ if $g_1, g_2$ are even and $\eta(\mathbf{g}) = -1$ otherwise.

For the effective Hamitonian $\mathcal{H}_{\text{eff}}$ in main text where the magnetic field is non-uniform but periodic and encloses one flux quantum per unit cell, the magnetic translation operator can be defined as above. We also separate the effective magnetic field into an average value, $\mathbf{B}(\mathbf{r}) = -B_0\hat{z}$, and a position-dependent part, denoted by $\delta\mathbf{B}(\mathbf{r}) = -\delta B(\mathbf{r})\hat{z}$, which has zero average. The corresponding vector potential can be split in a similar way so that
\begin{equation}
	\mathbf{B}^{\text{eff}}(\mathbf{r}) = -(B_0 + \delta B(\mathbf{r}))\hat{z} = \nabla\times \mathbf{A}_0(\mathbf{r}) + \nabla \times\delta\mathbf{A}(\mathbf{r})
\end{equation}
$\mathbf{A}_0$ is a linear function of position while $\delta\mathbf{A}(\mathbf{r})$ has the moir\'e superlattice periodicity. So the effective Hamiltonian can be written as
\begin{equation}\label{eq17}
	\begin{aligned}
		\hat{H} =& \frac{(\mathbf{p} - e\mathbf{A})^2}{2m} + \sum_{j=x,y}\frac{\hbar^2}{8m}(\partial_j\hat{\mathbf{S}})^2\\
		=& \omega_c \left[a^\dagger a + \frac{1}{2} + \frac{\delta A_{+}\delta A_{-}}{(lB_0)^2} + \frac{\delta A_{-}}{lB_0}a + a^\dagger \frac{\delta A_{+}}{lB_0} + \frac{1}{2}\frac{\delta B}{B_0} + \frac{1}{8eB_0}(\partial_j\hat{\mathbf{S}})^2\right]
	\end{aligned}
\end{equation}
where $\delta A_{\pm} = \frac{1}{\sqrt{2}}(\delta A_x \pm i \delta A_y)$. Fourier expanding the $\delta B(\mathbf{r})$
\begin{equation}
	\delta B(\mathbf{r}) = \sum_{\mathbf{g}}\delta B_{\mathbf{g}}e^{i\mathbf{g}\cdot\mathbf{r}}
\end{equation}
and use $-\delta B_\mathbf{g} = i\mathbf{g}\times\delta\mathbf{A}_\mathbf{g}$, we have the Fourier expansion of $\delta \mathbf{A}(\mathbf{r})$
\begin{equation}
	\delta\mathbf{A}(\mathbf{r}) = \sum_\mathbf{g}\mathbf{A}_\mathbf{g}e^{i\mathbf{g}\cdot\mathbf{r}} = i\sum_\mathbf{g}(g_y, -g_x)\frac{-\delta B_g}{|\mathbf{g}|^2}e^{i\mathbf{g}\cdot\mathbf{r}}.
\end{equation}
Besides, the Fourier expansion of $(\partial_j\hat{\mathbf{S}})^2$ can also be obtained numerically. Hence, the Hamiltonian matrix elements can be calculated by substituting the Fourier series of $\delta A_{+}, \delta A_{-}, \delta B, (\partial_j\hat{\mathbf{S}})^2$ into Eq.~\eqref{eq17} and utilizing Eq.~\eqref{eq14}. Diagonlizing the Hamiltonian matrix, the eigenstates can be written as $|i,\mathbf{k}\rangle = \sum_{n}z_{in\mathbf{k}}|n,\mathbf{k}\rangle$ where $i$ is band index.

Since the effective Hamiltonian is now expanded with LL magnetic Bloch basis, the overlap between i\textit{th} band in continuum model and n\textit{th} Landau level can be expressed as
\begin{equation}
	W_i = \frac{1}{N_{\text{uc}}}\sum_\mathbf{k} |z_{in\mathbf{k}}|^2
\end{equation}

From our calculation, the wave function overlap between the lowest two bands and the corresponding Landau levels are $99.74\%$ and $99.06\%$, respectively, which are almost the same.

\section{Fubini-Study Metric in Terms of Projectors}
In this section, we derive the numerical formula for Fubini-Study metric $g(\mathbf{k})$. It has roughly the same computational complexity with Wilson loop, which requires the projectors $P(\mathbf{k})$ defined on a fine mesh over the Brillouin zone (BZ). The following derivation follows the Appendix B of Ref.~\cite{Jonah2022}.

\subsection{Definition and numerical process}
$h(\mathbf{k})$ is a Hermitian Hamiltonian, which has the following spectral decomposition
\begin{equation}
h_{\alpha \beta}(\mathbf{k}) = \sum_n U_{\alpha n}(\mathbf{k}) E_n(\mathbf{k}) U^*_{\beta,n}(\mathbf{k})
\end{equation}
here $U_n(\mathbf{k})$ is a column eigenvector of $h(\mathbf{k})$ with eigenvalue $E_n(\mathbf{k})$. The spectrum $E_n(\mathbf{k})$ is named as the band structure. By defining the energy eigenstates through a unitary transition $\gamma^\dag_{\mathbf{k},n} = c^\dag_{\mathbf{k},\alpha} U_{\alpha,n}(\mathbf{k})$, the Hamiltonian can be diagonalized:
\begin{equation}
H = \sum_{\mathbf{k},n} E_n(\mathbf{k}) \gamma^\dag_{\mathbf{k},n} \gamma_{\mathbf{k},n} \\
\end{equation}
here we say that the Hamiltonian has a gap at filling $N_{\mathrm{occ}}$ if $E_{N_{\mathrm{occ}}+1}(\mathbf{k}) - E_{N_{\mathrm{occ}}}(\mathbf{k}) > 0$ for all $\mathbf{k}$. States in the occupied bands are all filled in the many-body ground state of the Hamiltonian. For ease of notation, we define $U(\mathbf{k}) = [U_1(\mathbf{k}), \dots, U_{N_{\mathrm{occ}}}(\mathbf{k})]$ as the $N_{\mathrm{orb}} \times N_{\mathrm{occ}}$ matrix of occupied eigenvectors. Because of the orthonormality of the eigenvectors, $U(\mathbf{k})$ has these important properties
\begin{equation}
U^\dag(\mathbf{k}) U(\mathbf{k}) = \mathbbm{1}_{N_{\mathrm{occ}} \times N_{\mathrm{occ}}}, \qquad U(\mathbf{k}) U^\dag(\mathbf{k}) \equiv P(\mathbf{k})
\end{equation}
here $P(\mathbf{k})$ is a Hermitian projector onto the occupied bands, which satisfies $P(\mathbf{k})^2 = P(\mathbf{k})$ and $P(\mathbf{k}) U(\mathbf{k}) = U(\mathbf{k})$. $U(\mathbf{k})$ is only defined up to $U(\mathbf{k}) \to U(\mathbf{k}) \mathcal{W}(\mathbf{k})$ where $\mathcal{W}(\mathbf{k})$ is an arbitrary $N_{\mathrm{occ}} \times N_{\mathrm{occ}}$ unitary matrix. The projector $P(\mathbf{k})$ is a gauge-invariant object on the BZ. We will now derive another gauge-invariant object, the abelian quantum geometric tensor $\eta_{\mu\nu}$, with two derivatives. In two dimension, a two derivative tensor is fundamental because it is dimensionless when integrated on BZ. In fact, we will see that the Berry curvature and Fubini-Study metric appear in this construction.

The starting point is to define $(1-U(\mathbf{k}) U^\dag(\mathbf{k})) \partial_i U(\mathbf{k})$. Under an arbitrary gauge transformation of the $N_{\text{occ}}$ eigenvectors $U \to U \mathcal{W}$, we find
\begin{equation}
  \begin{aligned}
(1-U U^\dag) \partial_\nu U \to (1-U U^\dag) \partial_\nu (U \mathcal{W}) &=  (1-U U^\dag) (\partial_\nu U \mathcal{W} + U  \partial_\nu \mathcal{W}) \\
&= \big( (1-U U^\dag) \partial_\nu U \big)\, \mathcal{W} \\
  \end{aligned}
\end{equation}
here in the last line we used the projector property $(1-U U^\dag) U = U (1- U^\dag U) = 0$. Note that the simple partial derivative $\partial_i U$ does not transform covariantly. Now we can define two-derivative a gauge-covariant $N_{\mathrm{occ}} \times N_{\mathrm{occ}}$ matrix called the quantum geometric tensor:
\begin{equation}
\eta_{\mu\nu}(\mathbf{k}) = \partial_\mu U^\dag (1 - U U^\dag) \partial_\nu U
\end{equation}
which also can be written as $\eta_{\mu\nu}  = \big( (1-U U^\dag) \partial_\mu U \big)^\dag (1-U U^\dag) \partial_\nu U$ because $(1-U U^\dag)^2 = (1-U U^\dag)$. Since $\eta_{\mu\nu}\to \mathcal{W}^\dag \eta_{\mu\nu} \mathcal{W}$ under gauge transformations, and we find that the \emph{abelian} quantum geometric tensor defined by
\begin{equation}
\text{Tr} \ \eta_{\mu\nu}(\mathbf{k}) \to \text{Tr} \ \mathcal{W}^\dag(\mathbf{k}) \eta_{\mu\nu}(\mathbf{k}) \mathcal{W}(\mathbf{k}) = \text{Tr} \ \eta_{\mu\nu}(\mathbf{k})
\end{equation}
is gauge invariant.
It is desirable to have expressions for the abelian quantum geometric tensor in terms of projector matrix which is the simple gauge-invariant object. We need the identity $(\del_\nu U^\dag) U = - U^\dag \del_\nu U$ which follows from $\del_\nu (U^\dag U) = \del_\nu \mathbbm{1} = 0$. With this, a direct calculation tells us that
\bea
\del_\nu P &= \del_\nu U \, U^\dag + U \del_\nu U^\dag \\
(\del_\nu P) U&= \del_\nu U  + U (\del_\nu U^\dag) U = \del_\nu U  - U U^\dag \del_\nu U = (1- U U^\dag) \del_\nu U \\
\eea
From it follows that $\eta_{\mu\nu} = U^\dag \del_\mu P  \del_\nu P \, U $ and thus
\bea
\Tr \eta_{\mu\nu} &= \Tr U^\dag \del_\mu P  \del_\nu P \, U =  \Tr U  U^\dag \del_\mu P  \del_\nu P = \Tr P \del_\mu P  \del_\nu P \
\eea
from the cyclicity of the trace. This expression for $\Tr \eta_{\mu\nu}$ is obviously gauge invariant. Note that $\Tr \eta^\dag_{\mu\nu} =  \Tr \del_\nu P  \del_\mu P  P = \Tr \eta_{\mu\nu}$ so the symmetric part of $\Tr \eta_{\mu\nu}$ is real and the anti-symmetric part is imaginary. We now can separate $\Tr \eta_{\mu\nu}$ into symmetric and anti-symmetric parts:
\bea
\Tr \eta_{\mu\nu} &= g_{\mu\nu} - \frac{i}{2} f_{\mu\nu}, \qquad g_{\mu\nu} =  \frac{1}{2} \Tr P \{\del_\mu P, \del_\nu P \}, \quad f_{\mu\nu} = i \Tr P [\del_\mu P, \del_\nu P] \\
\eea
where $g_{\mu\nu}$ is the Fubini-Study metric and $f_{\mu\nu}$ is the Berry curvature. For the Fubini-Study metric, there is a simpler expression using the identities $\{P,\del_\nu P\} = \del_\nu P$ (derived from $P^2 = P$) and $\Tr A \{B,C\} = \Tr B \{A,C\}$, from which we can find
\bea
g_{\mu\nu} &= \frac{1}{2} \Tr P \{\del_\mu P, \del_\nu P \} = \frac{1}{2} \Tr \del_\mu P \del_\nu P \ .
\eea

Thus we derive a simple point-split expression for the numerical calculation of the diagonal term of $g_{\mu\nu}$.
\bea
\frac{1}{2} \Tr (\del_{k_x} P)^2 &= \frac{1}{2\eps^2} \Tr \lp P(\mbf{k} + \eps \hat{x}) - P(\mbf{k}) \rp^2 \\
&= \frac{1}{2\eps^2} \Tr \lp P(\mbf{k} + \eps \hat{x})^2 + P(\mbf{k})^2 - \{P(\mbf{k} + \eps \hat{x}) ,P(\mbf{k})  \}\rp \\
&= \frac{1}{2\eps^2} \Tr \lp P(\mbf{k} + \eps \hat{x}) + P(\mbf{k}) - \{P(\mbf{k} + \eps \hat{x}) ,P(\mbf{k})  \}\rp \\
&= \frac{1}{2\eps^2} \lp 2N_{\mathrm{occ}} - 2\Tr P(\mbf{k} + \eps \hat{x}) P(\mbf{k}) \rp \\
&= \frac{1}{\eps^2} \lp N_{\mathrm{occ}} - \Tr P(\mbf{k} + \eps \hat{x}) P(\mbf{k}) \rp \\
\eea
where we used $P(\mbf{k})^2 = P(\mbf{k})$ and $\Tr P(\mbf{k}) = N_{\mathrm{occ}}$.

\section{Extended data for exact diagonalization}
\subsection{Cluster geometry}
\label{Cluster_Geometry}
 We defined the cluster geometry used for finite-size exact diagonalization calculation in main text and this supplementary materials.
 
\begin{figure}[htb]
  \begin{center}
    \includegraphics[width=1.0\columnwidth, clip=true]{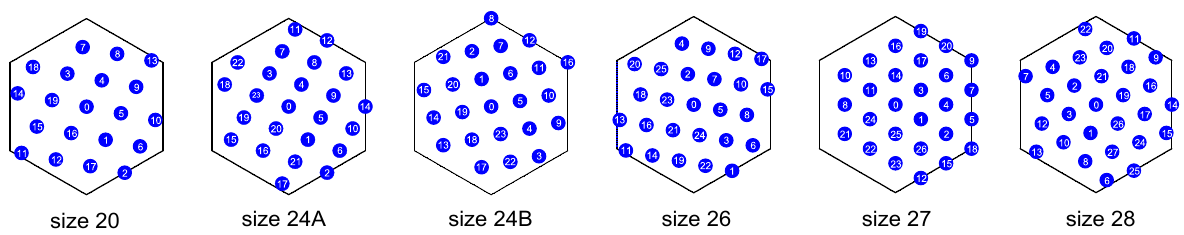}
  \end{center}
  \caption{Momentum space mesh diagrams for different cluster geometries used in this work.}
  \label{figS6}
\end{figure}

 These torus clusters are defined by two vectors $\mathbf{L}_i$ so that the systems have periodic boundary conditions $\mathbf{r} = \mathbf{r} + \mathbf{L}_i$. Then all allowed crystal momenta are $\mathbf{k} = N_1 \mathbf{T}_1 + N_2 \mathbf{T}_2$ where $\mathbf{T}_i = 2\pi \epsilon_{ij}\mathbf{L}_i\times \hat{z}/|\mathbf{L}_1\times\mathbf{L}_2|$ and $\mathbf{T}_i$ satisfy $\mathbf{T}_i\cdot\mathbf{L}_j = 2\pi \delta_{ij}$. Different cluster geometries are constructed by selecting $(\mathbf{L}_1^{(20)}, \mathbf{L}_2^{(20)}) = ((4, 0), (0, 5))$, $(\mathbf{L}_1^{(24A)}, \mathbf{L}_2^{(24A)}) = ((4, 0), (0, 6))$, $(\mathbf{L}_1^{(24B)}, \mathbf{L}_2^{(24B)}) = ((6, 0), (-1, -4))$$(\mathbf{L}_1^{(26)}, \mathbf{L}_2^{(26)}) = ((2, 4), (0, 13))$, $(\mathbf{L}_1^{(27)}, \mathbf{L}_2^{(27)}) = ((6, -3), (-3, 6))$, $(\mathbf{L}_1^{(28)}, \mathbf{L}_2^{(28)}) = ((4, 2), (-2, 6))$ for size $N = 20, 24, 24, 26, 27, 28$ respectively and parenthesis $(m, n)$ denotes $\mathbf{L}_i = m \mathbf{a}_1 + n \mathbf{a}_2$ with $\mathbf{a}_1 = a_0 (\frac{\sqrt{3}}{2}, -\frac{1}{2})$ and $\mathbf{a}_2 = a_0 (\frac{\sqrt{3}}{2}, \frac{1}{2})$. We show the momentum space mesh generated by $
 \mathbf{T}_1,\mathbf{T}_2$ for different sizes in Fig.~\ref{figS6}.

\section{Effective magnetic field}
\begin{figure}[htb]
	\begin{center}
		\includegraphics[width=0.4\columnwidth, clip=true]{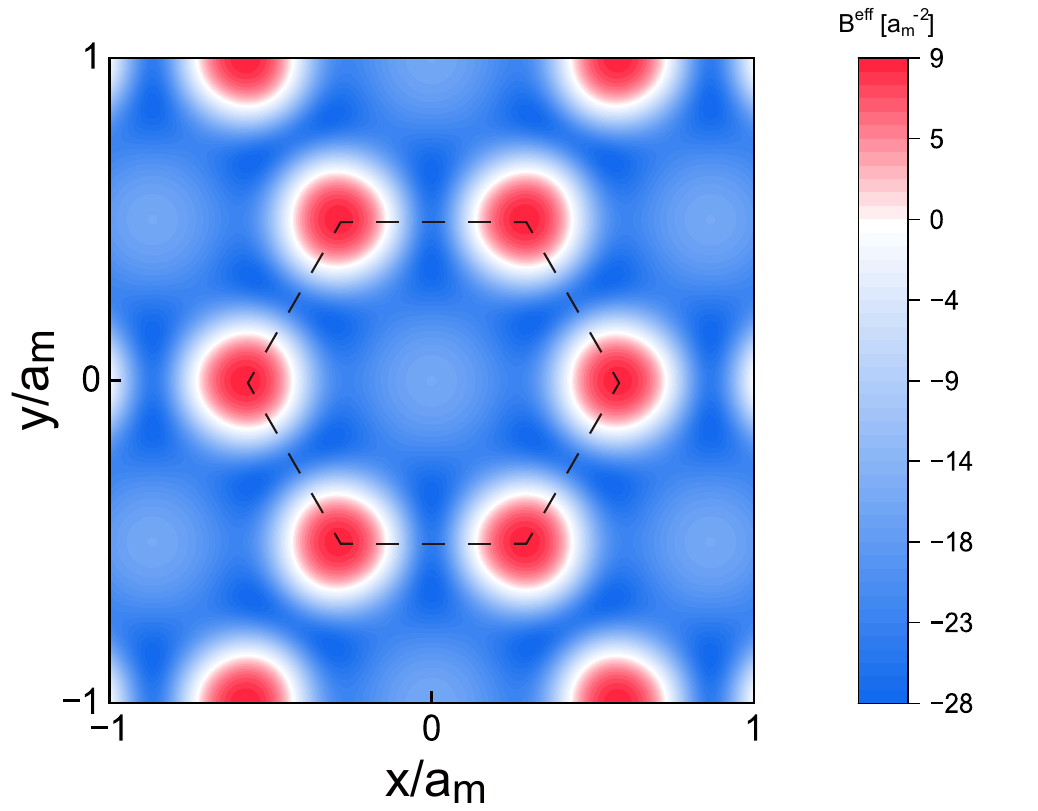}
	\end{center}
	\caption{Effective magnetic field emerges from noncoplanar periodic moir\'e potential (parameters are selected as $\alpha=1, N_0=0.28$).}
	\label{figS7}
\end{figure}

As mentioned in main text, the noncoplanar skyrmion texture will induce an effective magnetic field, which can be expressed as
\begin{equation}
	\mathbf{B}^{\text{eff}}(\mathbf{r}) = \frac{\hbar}{2e}\hat{\mathbf{S}}\cdot (\partial_x \hat{\mathbf{S}}\times \partial_y \hat{\mathbf{S}})
\end{equation}
where $\hat{\mathbf{S}}(\mathbf{r})$ is periodic moir\'e potential defined in main text. The distribution of $\mathbf{B}^{\text{eff}}(\mathbf{r})$ is shown in Fig.~\ref{figS7}. Then the flux per primitive unit cell $\Phi$ can be calculated and $\Phi = -1$ in unit of flux quantum. 

%